\documentclass[preprint,3p]{elsarticle}

\usepackage[utf8]{inputenc}

\usepackage{graphicx}
\usepackage{hhline}
\usepackage{amsmath,amssymb}
\usepackage[version=4]{mhchem}
\usepackage{siunitx}
\usepackage{bbold}
\usepackage{ulem}
\usepackage{mathrsfs}
\usepackage[T1]{fontenc}
\usepackage{textcomp}
\usepackage{lmodern}
\usepackage{xcolor}
\usepackage{relsize}
\usepackage[colorlinks=true, allcolors=blue]{hyperref}
\usepackage{longtable,tabularx}
\usepackage[font=small]{caption}
\usepackage{tikz}
\usepackage{nicefrac}
\usepackage{array}
\usepackage{algorithm}
\usepackage{algpseudocode}
\usetikzlibrary{shapes.geometric, arrows}
\usetikzlibrary {arrows.meta}
\tikzstyle{startstop} = [rectangle, rounded corners, minimum width=3cm, minimum height=1cm,text centered, draw=black, text width=3.75cm, fill=red!30]
\tikzstyle{io} = [trapezium, trapezium left angle=70, trapezium right angle=110, minimum width=3cm, minimum height=1cm, text centered, draw=black, fill=blue!30]
\tikzstyle{process} = [rectangle, minimum width=3cm, minimum height=1cm, text centered, draw=black, fill=orange!30]
\tikzstyle{decision} = [diamond, minimum width=3cm, minimum height=1cm, text centered, draw=black, fill=green!30]
\tikzstyle{arrow} = [thick,->,>=stealth]
\tikzstyle{flow} = [circle, minimum width=1cm, minimum height=1cm, text centered, draw=black, fill=green!30]
\tikzstyle{AA} = [rectangle, rounded corners, minimum width=0cm, minimum height=1cm,text centered, draw=black, text width=2.cm, fill=black!30]
\tikzstyle{arrow} = [thick,->,>=stealth]

\newcommand{\LO}{{\textrm{LO}}}
\newcommand{\HO}{{\textrm{HO}}}
\newcommand{\thh}{{\textrm{th}}}
\newcommand{\hlf}{\textrm{\textonehalf}}
\newcommand{\e}{\textrm{e}}
\newcommand{\cc}{\textrm{c}}
\newcommand{\T}{\textrm{t}}
\newcommand{\p}{\textrm{p}}
\newcommand{\ii}{\textrm{i}}
\newcommand{\ppc}{\mathrm{ppc}}

\journal{Journal of Computational Physics}
\begin{document}
\begin{frontmatter}
\title{A High-Order Low-Order extended moment method for the Vlasov-Darwin particle-in-cell system}

\author[1]{Derek A. Kuldinow\fnref{fn1}\corref{cor1}}\ead{kuldinow@stanford.edu}
\author[2]{William T. Taitano\fnref{fn2}}
\author[1]{Kentaro Hara\fnref{fn3}}
\affiliation[1]{organization={Department of Aeronautics and Astronautics, Stanford University},addressline = {496 Lomita Mall},city={Stanford},state={California},postcode={94305},country={USA}}
\affiliation[2]{organization={Theoretical Division, Los Alamost National Laboratory},city={Los Alamos},state={New Mexico},postcode={87545},country={USA}}
\cortext[cor1]{Corresponding Author}
\fntext[fn1]{Ph.D. Candidate, Stanford University}
\fntext[fn2]{Research and Development Scientist, Los Alamos National Laboratory}
\fntext[fn3]{Assistant Professor, Stanford University}

\begin{abstract}
In this study, we develop an extended implicit moment method, namely, a coupled high-order low-order~(HOLO) method and apply it to the electromagnetic Vlasov-Darwin model. The high-order~(HO) system evolves particles in a manner that conserves charge, energy, and canonical momentum, while the low-order~(LO) system solves the fluid moment and Darwin equations, acting as an algorithmic convergence accelerator to the HO system. We demonstrate the HOLO method's ability to take timesteps far larger than the explicit limit, and accurately recover the system's evolution so long as its dynamical timescale is respected. Also, we find that the choice of LO fluid moment equations has a strong impact on the nonlinear convergence of the coupled particle-field system. The HOLO algorithm is benchmarked against electrostatic Landau damping and the electromagnetic electron and ion Weibel instabilities.
\end{abstract}

\begin{keyword}
    Implicit moment methods \sep
    Energy-conserving implicit PIC \sep
    Vlasov-Darwin model \sep
    Electromagnetic PIC \sep
    HOLO schemes
\end{keyword}
\end{frontmatter}

\section{Introduction}
\vspace{10pt}

The Vlasov-Maxwell system of equations describes the time evolution of a collisionless plasma and its induced electromagnetic fields. The particle-in-cell (PIC) method~\cite{birdsall2018plasma, hockney2021computer, grigoryev2012numerical} solves the Vlasov equation by approximating the plasma velocity distribution function~(VDF) as a collection of many particles, which are then evolved using the equations of motion. The evolution of the particles is coupled to the electromagnetic fields through the moments of the VDF, i.e., the charge and current densities. Likewise, the fields are coupled to the particles' motion through the Lorentz force. An efficient and accurate solution of the coupled Vlasov-Maxwell system is key to predictive modeling of plasmas. In this study, we develop an extended implicit moment method for charge- and energy-conserving PIC simulations of nonrelativistic electromagnetic plasmas and characterize the performance of the methods.\par

Explicit PIC methods are attractive because they are often highly parallelizable, straightforward to implement and, importantly, inexpensive per timestep. For these reasons, explicit schemes have been successfully applied to a wide variety of plasma applications~\cite{birdsall2018plasma, bowers2008ultrahigh, bowers2009advances}. However, standard explicit PIC schemes are constrained by a number of criteria for numerical stability. The presence of the finite grid instability typically requires that the mesh resolve the Debye length, $\Delta x<\lambda_{\rm D} = \sqrt{\epsilon_0k_{\rm B}T_\e/\left(ne^2\right)}$; likewise, temporal stability requires that the timestepping resolve the plasma frequency, $\Delta t<\omega_{\p\e}^{-1}=\sqrt{m_\e\epsilon_0/\left(ne^2\right)}$, where $\Delta x$ and $\Delta t$ are the grid size and timestep, respectively, $\epsilon_0$ is the permittivity of free space, $k_{\rm B}$ is Boltzmann's constant, $T_\e$ is the electron temperature, $n$ is the quasineutral plasma density, $e$ is the fundamental charge, and $m_\e$ is the electron mass. For electromagnetic systems, another timestep restriction is the Courant-Friedrichs-Lewy~(CFL) condition based on the speed of light, $\Delta t<\Delta x/c$, where $c$ is the speed of light. In multiple spatial dimensions, the CFL condition must be satisfied in all dimensions, contributing to a restrictive timestep that can be orders of magnitude smaller than the system's dynamical timescales.\par

Implicit PIC methods, in contrast, iteratively converge the field and particle updates, significantly mitigating finite-grid instabilities~\cite{barnes2021finite} and enabling the use of larger timesteps comparable to the dynamical timescales of interest. However, the efficiency of these methods critically depends on the convergence behavior of the iterative coupling between the particle motion and the field evolution. For especially large timesteps, direct-implicit methods (DIMs)~\cite{friedman1981direct,langdon1983direct}, which tightly couple the particles to the fields, require solving a high-dimensional system that includes the degrees of freedom of both the particles and the fields. Since storing the full particle system during iterative solves is memory-intensive, prior studies~\cite{chen2011energy,markidis2011energy,chen2014energy} have employed a particle-enslavement strategy. In this approach, particle equations of motion are evolved as a function evaluation within the electromagnetic field residuals of a Newton-Krylov solver, thereby reducing memory demands while maintaining implicit coupling.\par

Another approach to addressing the challenges posed by stiff plasma timescales is the class of implicit moment methods (IMMs)~\cite{denavit1981time, mason1981implicit, brackbill1982implicit, brackbill1985simulation}. IMMs offer an alternative to DIMs by introducing auxiliary fluid moment equations, often closed at the stress-tensor level obtained self-consistently from particle information. These fluid equations are then coupled with the field equations to provide an implicit estimate of the system state, thereby alleviating stiffness associated with fast plasma waves. However, traditional IMMs do not enforce strict consistency between the moment and particle solutions, causing them to suffer from mismatches between the particle evolution and the moment-field subsystem, which can lead to nonphysical or unrealizable results. High-Order Low-Order (HOLO) methods~\cite{taitano2013development,chacon2017multiscale} extend the IMM framework by introducing an implicit iteration between the high-order (HO) particles and the low-order (LO) moment-field auxiliary system. This iterative procedure incorporates so-called \textit{consistency terms}, which ensures strict convergence to a consistent solution across both systems.\par

In addition to addressing stiff electromagnetic timescales and ensuring overall system stability, numerical solutions of the Vlasov-Maxwell system should respect the system’s inherent conservation laws. Key among these are the conservation of charge, momentum, and energy. Standard PIC algorithms~\cite{birdsall2018plasma} typically conserve only charge and momentum exactly, while others~\cite{lewis1970energy} conserve charge and energy only in the asymptotic limit as $\Delta t \rightarrow 0$. Failure to conserve energy exactly can result in spurious plasma heating or cooling, potentially yielding physically unrealizable solutions. To address these issues, implicit and structure-preserving PIC schemes~\cite{chen2011energy,  markidis2011energy,chen2014energy,kraus2017_gempic,morrison2017structure,jianyuan2018structure} have been developed to rigorously enforce or systematically bound conservation errors. Additionally, recent advances in explicit PIC formulations introduce Lagrange multiplier-like terms to ensure exact energy conservation~\cite{ricketson2005energyconspic}. \par

In this study, we extend the HOLO electrostatic PIC method proposed in Ref.~\cite{taitano2013development} to one-dimensional \textit{electromagnetic} plasmas and study the effects of the choice of LO moment system on the nonlinear convergence. We adopt the Darwin equations~\cite{chen2014energy} as an alternative to the full Maxwell equations due to their validity in the non-relativistic regime and their elimination of the local speed-of-light CFL constraint. Our approach couples the implicit, \mbox{charge-,} \mbox{canonical-momentum-,} and energy-conserving electromagnetic particle pusher (the HO system), as described in Ref.~\citenum{chen2014energy}, with a fluid moment-Darwin system (the LO system). For the fluid moment equations, we compare the algorithmic performance of the 4-moment~(continuity and 3 momenta), 5-moment~(4-moment system plus normal stress), and 7-moment~(5-moment system plus shear stress) systems for a one-dimensional plasma. We also highlight the ability of the HOLO PIC method to stably take timestep sizes orders of magnitude larger than those permitted by explicit methods. \par

The rest of this paper is organized as follows. In \S\ref{sec:VD} we present the continuum form of the governing Vlasov-Darwin equations in one spatial dimension and three velocity dimensions (1D-3V). In \S\ref{sec:HOLO}, we provide an overview of the HOLO solver and describe the LO fluid moment systems used to solve for the fields. We also discuss the justification for the advantages offered by different choices of LO systems through an analysis of their dispersion relations. The spatial and temporal discretization of the equations is presented in \S\ref{sec:Discretization}, as well as a description of the overall algorithm, including how the equations are solved. In \S\ref{sec:Tests}, we present numerical results and solver statistics for electrostatic and electromagnetic test cases, comparing the choices for the LO system and timestep size, followed by concluding remarks in \S\ref{sec:Conclusions}.

\section{The Vlasov-Darwin System}\label{sec:VD}
\vspace{10pt}

The Darwin model~\cite{hockney2021computer,darwin1920li} is a non-relativistic approximation to order $v^2/c^2$ of Maxwell's equations (here $v$ is the particle velocity), which eliminates the propagation of light waves that have been seen to cause spurious oscillations born from particle noise and numerical Cherenkov radiation~\cite{langdon1979analysis}. The Darwin system has been used in a number of previous studies of implicit electromagnetic PIC simulations~\cite{chen2014energy,hewett1985elimination,nielson1976particle} and fluid simulations~\cite{kuldinow2025ten} due to its favorable numerical properties for large timestep sizes. The Vlasov-Darwin equations for the time evolution of a collisionless electromagnetic plasma can be written as~\cite{chen2014energy,nielson1976particle}
\begin{equation}\label{eq:Vlasov}
    \partial_tf_s+v_i\partial_i f_s+\frac{q_s}{m_s}\left(E_i+\varepsilon_{ijk}v_jB_k\right)\partial_{v_i}f_s=0,
\end{equation}
\begin{equation}\label{eq:Ampere}
    \partial_k\partial_kA_i = -\mu_0j_i+\mu_0\epsilon_0\partial_t\partial_i\phi,
\end{equation}
\begin{equation}\label{eq:Gauss}
    \epsilon_0\partial_k\partial_k\phi=-\rho,
\end{equation}
\begin{equation}\label{eq:gradA}
    \partial_kA_k=0,
\end{equation}
where $f_s = f_s\left(x_i,v_i,t\right)$ is the velocity distribution function (VDF) as a function of space, velocity, and time, respectively, of species $s$ with electric charge $q_s$ and mass $m_s$; $\partial_t = \partial/\partial t$, $\partial_i = \partial/\partial x_i$, and $\partial_{v_i}=\partial/\partial v_i$; $E_i$, $B_i$, $\phi$ and $A_i$ are the electric field, magnetic field, scalar electrostatic potential and magnetic vector potential, respectively; $\varepsilon_{ijk}$ is the Levi-Civita tensor, $\rho$ and $j_i$ are the charge and current densities, respectively, $\mu_0$ is the  permeability of free space, and subscripts $i$, $j$, $k$ denote Einstein summation notation. Unlike Maxwell's equations, only the Coulomb gauge is physically consistent for the Darwin system, as described in Refs.~\citenum{chen2014energy} and \citenum{darwin1920li}. The charge and current densities are calculated from the distribution function as
\begin{equation}\label{eq:rhoj}
    \rho = \sum_s q_s\int
    f_sd^3v, \hspace{15pt}
    j_i = \sum_s q_s\int
    v_if_sd^3v, 
\end{equation}
where the integral is over all of velocity space. The electric and magnetic fields are found from the potentials as
\begin{equation}\label{eq:EB}
    E_i = -\partial_i\phi-\partial_tA_i,
    \hspace{15pt}
    B_i = B_{0,i}+\varepsilon_{ijk}\partial_jA_k,
\end{equation}
where $B_{0,i}$ is an applied external magnetic field. The Vlasov equation (Eqn.~\eqref{eq:Vlasov}) describes how the electric and magnetic fields are coupled to the particle dynamics, while the Darwin equations (Eqns.~\eqref{eq:Ampere}-\eqref{eq:gradA}) describe how the particles are coupled to the electromagnetic potentials through the moments of the VDF. \par

In one spatial dimension, $x$, the Darwin equations (Eqns. \eqref{eq:Ampere}-\eqref{eq:gradA}) can be written as
\begin{equation}\label{eq:ddxAx}
    \partial_x^2A_x = -\mu_0j_x+\mu_0\epsilon_0\partial_t\partial_x\phi,
\end{equation}
\begin{equation}\label{eq:ddxAperp}
    \partial_x^2A_\perp = -\mu_0j_\perp,
\end{equation}
\begin{equation}\label{eq:ddxphi}
    \epsilon_0\partial_x^2\phi = -\rho,
\end{equation}
\begin{equation}\label{eq:dxAx}
    \partial_xA_x = 0,
\end{equation}
where $\perp \in \{y,z\}$. Equation~\eqref{eq:dxAx} implies that $A_x$ is constant over the domain and does not change in time; combining this with Eqn.~\eqref{eq:ddxAx} and taking the spatial derivative implies that
\begin{equation}
    \epsilon_0\partial_t\left(\partial_x^2\phi\right) = \partial_x j_x,
\end{equation}
which, alongside Eqn.~\eqref{eq:ddxphi}, necessitates local charge conservation: $\partial_t\rho = -\partial_x j_x$. Since $A_x$ is constant over the domain and in time, $E_x = -\partial_x \phi$ from Eqn.~\eqref{eq:EB}. \par 
Then, the final set of one-dimensional Darwin equations can be written from Eqs.~\eqref{eq:EB}, \eqref{eq:ddxAx}, and \eqref{eq:ddxAperp},  as 
\begin{equation}\label{eq:dtEx}
    \epsilon_0\partial_tE_x + j_x-\langle j_x\rangle = 0,
\end{equation}
\begin{equation}\label{eq:ddxAperpper}
    \frac{1}{\mu_0}\partial_x^2A_\perp +j_\perp-\langle j_\perp\rangle=0,
\end{equation}
\begin{equation}\label{eq:Bi} 
    B_i = B_{0,i}+\varepsilon_{ijk}\partial_jA_k,\\
\end{equation}
\begin{equation}\label{eq:Eperp} 
    E_\perp = -\partial_tA_\perp,\\
\end{equation}
where $\langle\cdot\rangle=\frac{1}{L}\int_0^L\left(\cdot\right) dx
$ denotes the spatial average over a periodic domain of length $L$ and is included to preserve Galilean invariance in a periodic domain~\cite{chen2011energy,hasegawa1968one}.
Note that the scalar potential is not explicitly solved for; rather, so long as Gauss's law, Eqn.~\eqref{eq:ddxphi}, is satisfied for the initial conditions, it is satisfied for all time given (local) conservation of charge and Eqn.~\eqref{eq:dtEx}.\par
The Darwin equations then coupled to the Vlasov equation (Eqn.~\eqref{eq:Vlasov}). In a 1D-3V PIC simulation, the velocity distribution function $f$ is assumed to be represented by a collection of (macro)particles:
$$f(x,v_i,t) = \sum_pw_p\delta\left(x-x_p(t)\right)\delta^3\left(v_i-v_{p,i}(t)\right),$$ where $w_p$ is the weight of the particle $p$ and $\delta$ is the Dirac delta function. Substituting this form of $f$ back into Eqn.~\eqref{eq:Vlasov} yields the equations of motion for each particle $p$:
\begin{align}
        \partial_tx_p &= v_{p,x},\label{eq:EoM1}
        \\
        \partial_t v_{p,i} &= \frac{q_s}{m_s}\left[E_{p,i}+\varepsilon_{ijk}v_{p,j}B_{p,k}\right],\label{eq:EoM2}
\end{align}
where $E_{p,i}=E_i\left(x_p\right)$, and $B_{p,k}=B_k\left(x_p\right)$ represent the electric and magnetic fields, respectively, at the particle position.\par

Due to its elliptical nature, the Darwin system is unconditionally unstable for domain sizes greater than the electron skin depth $(c/\omega_{\p\e})$ with explicit time integration; furthermore, as discussed above, classic explicit PIC schemes are limited to resolving the electron plasma frequency and Debye length. However, dynamics of interest can occur on lab scales much larger than the skin depth and Debye length and timescales much longer than the inverse plasma frequency. For these reasons, we seek an implicit scheme which can take large timesteps and simulate large domain sizes without being prohibitively expensive.

\section{The HOLO Multiscale Solver}\label{sec:HOLO}
\vspace{10pt}

To solve the 1D-3V particle-kinetic Vlasov-Darwin equations (Eqns.~\eqref{eq:dtEx}-\eqref{eq:EoM2}), we use a coupled implicit high-order low-order~(HOLO) scheme. In the HOLO scheme, we employ a high degree-of-freedom particle time-integrator (the HO system) and a low degree-of-freedom fluid moment-based solver for the electromagnetic fields (the LO system). The LO system efficiently obtains an implicit update for the fields by capturing stiff coupling with the auxiliary fluid equations that are closed by particle moments. This effort expands on the previous \textit{electrostatic} work in Ref.~\citenum{taitano2013development} that pioneered the development of IMMs that enforce exact consistency between the HO and LO systems. In this section, we provide a general overview of the HOLO system before describing each part of the algorithm.\par

\begin{figure}[h!]
\centering
\begin{tikzpicture}[node distance=2cm]
\draw [thick,rounded corners] (5,-2.875) rectangle (-5.5,-6); 
\fill [blue!10!white,rounded corners] (4.5,-0.75) rectangle (-5,-5); 
\node (in1) [process,yshift=0.5cm] {Initial Conditions};
\node (HO) [startstop,below of = in1,yshift = -0.5cm] {Particle Push (HO) \smaller{ Picard Iteration}};
\node (LO) [startstop,below of = HO,xshift = 1cm] {Field Solve (LO) \smaller{JFNK Iteration}};
\node (And)[AA,left of = LO, xshift = -2.5cm]{Anderson\\mixing};
\node (HOLO) [flow,right of = HO,xshift=1.5cm,yshift=0cm]{$\mathcal{M}_\alpha,\gamma_\alpha$};
\node (LOHO) [flow,left of = HO,xshift=-1.5cm,yshift=0cm]{$E_i,A_i$}; 
\node[] at (-3.5,-1.1) {HOLO iteration};
\node (A) at (0,-0.9) {};
\draw [arrow] (in1) -- (A); 
\draw [arrow] (HO) -- (HOLO);
\draw [arrow] (HOLO) |- (LO);
\draw [arrow] (LO) -- (And);
\draw [arrow] (And) -- (LOHO);
\draw [arrow] (LOHO) -- (HO);
\node[] at (0,-5.5) {$\mathlarger{\Delta t}$};
\draw [-{Latex[length=5mm]}] (2,-6) -- (-0.25,-6);
\end{tikzpicture}
\caption{Flowchart of the HOLO scheme} \label{fig:HOLOflow}
\end{figure}
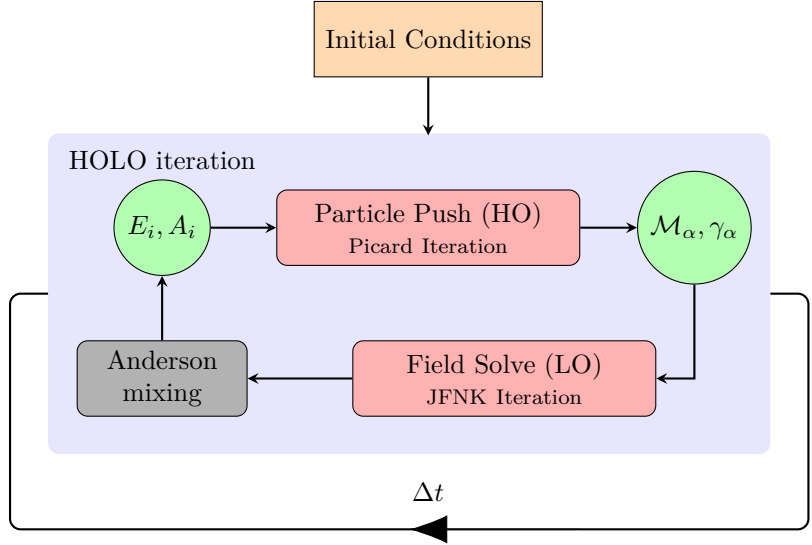

A schematic of the HOLO algorithm is provided in Figure~\ref{fig:HOLOflow}. Initial positions and velocities of the particles are specified by the user, as well as any applied magnetic field. Self-consistent induced electric and magnetic fields are calculated using Eqns.~\eqref{eq:ddxphi} and \eqref{eq:ddxAperpper}. The solutions at each timestep are then converged using the fully-implicit HOLO iteration, which consists of an HO step and an LO step.\par

\subsection{HO Solver}\label{sec:HOLO_HO}
For the HO system, we employ the implicit Crank-Nicolson discretization described in Refs.~\citenum{chen2011energy} and \citenum{chen2014energy} to solve Eqns.~\eqref{eq:EoM1}-\eqref{eq:EoM2} and advance particles each timestep. The electromagnetic fields are not evolved within the HO system. The full timestep $\Delta t$ is divided into smaller substeps $\Delta\tau$ to enforce exact local charge as well as to minimize numerical errors~\cite{chen2011energy}. Particle positions, velocities, and substep sizes are calculated using Picard iteration until all particles complete the full timestep. After convergence of the HO particle update, moments of the velocity distribution function, $\mathcal{M}_\alpha$, are computed at each grid point. A detailed description of the HO discretization is given in \S\ref{sec:HO_discrete}. By construction, the algorithm exactly conserves local charge and per-particle canonical momentum, while energy is conserved up to the tolerance defined by the HOLO iteration. \par

In order to ensure consistency between the HO and  LO equations, we also calculate so-called consistency terms $(\gamma_\alpha)$, which quantify the inconsistencies between the HO and LO systems, including errors due to the spatio-temporal discretization of the LO system and the use of a finite number of particles in the HO system. The consistency terms are described in \S\ref{sec:LO_discrete}. The moments and the consistency terms are then fed into the LO system. The converged estimates for the LO quantities including the fields are combined with the estimates from the previous iterations using Anderson mixing ~\cite{anderson1965iterative, walker2011anderson, willert2014leveraging} to ensure robust convergence of the method; the electromagnetic field estimates are then fed back into the HO system. \par 

\subsection{LO Systems}\label{sec:HOLO_LO}

The LO system solves for the electromagnetic fields at the next timestep by solving Eqns.~\eqref{eq:dtEx} and \eqref{eq:ddxAperpper} coupled with an auxiliary set of fluid equations (to be discussed shortly). However, as noted in previous work~\cite{denavit1981time,mason1981implicit,taitano2013development, willert2014leveraging}, if the current densities from the HO system are directly substituted into the Darwin system (i.e., the plasma fluid equations are not included in the LO scheme),
the implicit response of the electromagnetic fields is not effectively captured and the HOLO iteration is unstable for $\Delta t>\omega_{\p\e}^{-1}$. For this reason, the Darwin field equations are coupled to and solved in tandem with the plasma fluid equations, which results in an implicit update to the field that is more accurate, thus relaxing the restriction on the timestep and allowing for timestep sizes larger than $\omega_{\p\e}^{-1}$. The LO system is constructed by choosing the fluid equations to be solved and how the set of LO equations is closed with particle information from the HO system. As with any fluid moment system (e.g, $\left\{ \mathcal{M}_{0}, \mathcal{M}_{1}, \ldots, \mathcal{M}_{N} \right\}$, where $N$ is the number of moments accounted for), closure is required for the next highest moment that is not solved for (i.e., $\mathcal{M}_{N+1}$). For each LO system, we consider two types of closure: \textit{conservative} and \textit{primitive}. In the conservative closure, the closing moment, $\mathcal{M}_{N+1}$ is obtained from the HO system, while in the primitive closure, only the next highest \textit{central} moment is calculated from the HO system. Below we propose three LO systems: the 4-moment, 5-moment, and 7-moment systems.

\subsubsection{4-moment system}

Previous work with implicit moment methods has employed the 4-moment system~\cite{brackbill1985simulation,taitano2013development}, where the only fluid equations included are the continuity and momentum equations, solving for the evolution of the number density $n$ and momentum density $\Gamma$ of a species $s$:
\begin{equation}\label{eq:4M}
\begin{cases}
    \mathlarger{\partial_tn_{s}+\partial_x\Gamma_{s,x} = 0}\\
    \\
    \mathlarger{\partial_t\Gamma_{s,i}+\partial_xS_{s,ix} = a_{s,i},}
\end{cases}
\end{equation}
where $a_{s,i} = \frac{q_s}{m_s}\left(n_s E_i+\varepsilon_{ijk}\Gamma_{s,j} B_k\right)$ is the acceleration density, and $S_{s,ij}$ is the species momentum flux density, i.e., the stress tensor. Here, $E_i$ and $B_k$ are the electric and magnetic fields that are coupled with the Darwin equations in Eqns.~\eqref{eq:dtEx}-\eqref{eq:Eperp}.\par

In the conservative closure, $S_{ij} = n\tilde{S}_{ij}^\HO$, where $\tilde{S}_{ij}^\HO=\int v_i v_j\hat{f}d^3v$ is the density-normalized stress tensor calculated from the HO system and $\hat{f} = f/\int fd^3v$ is the normalized VDF. The momentum flux density can be written as $S_{ij} = n T_{ij}+\Gamma_i\Gamma_j/n$, where $T_{ij}=\int w_i w_j \hat{f} d^3 v$ is the anisotropic temperature tensor, $w_i = v_i - u_i$ is the peculiar velocity, and $u_i = \Gamma_i / n$ is the bulk velocity. In contrast, the primitive closure uses the \textit{central} moment from the HO system; hence, the primitive closure requires $T_{ix}^\HO=\tilde{S}_{ix}^\HO-\Gamma_x^\HO\Gamma_i^\HO/\left(n^\HO\right)^2$, where the superscript $\HO$ refers to the properties calculated from the HO system rather than solved for in the LO system. In summary, the equations for the 4-moment system can be written as 
\begin{equation}\label{eq:4MCP}
\textrm{Conservative: }
\begin{cases}
    \mathlarger{\partial_tn_{s}+\partial_x\Gamma_{s,x} = 0}\\
    \\
    \mathlarger{\partial_t\Gamma_{s,i}+\partial_x\left(n_s\tilde{S}^\HO_{s,xi}\right) = a_{s,i},}
\end{cases}
\textrm{Primitive: }
\begin{cases}
    \mathlarger{\partial_tn_{s}+\partial_x\Gamma_{s,x} = 0}\\
    \\
    \mathlarger{\partial_t\Gamma_{s,i}+\partial_x\left(n_s T^\HO_{s,xi}+\frac{\Gamma_{s,x}\Gamma_{s,i}}{n_s}\right) = a_{s,i}.}
\end{cases}
\end{equation}
To motivate comparing the two formulations, we analyze the difference in the dispersion relations of a simplified system that neglects the electromagnetic bulk forces and observe what waves are captured. Through linear analysis, i.e., considering a perturbation of the form $\exp[-i(\omega t-kx)]$, where $\omega$ is the frequency and $k$ is the wavenumber, we find the following dispersion relations:
\begin{equation}
\begin{array}{ll}
\label{eq:4M_Disp}
    \textrm{Conservative: } & v_w= \left\{0,\pm\sqrt{\tilde{S}_{xx}}\right\} = \left\{0,\pm\sqrt{u_x^2+v_\thh^2}\right\}, \\[5pt]
    
    \textrm{Primitive: } & v_w= \left\{u_x,u_x\pm v_\thh\right\},
\end{array}
\end{equation}
where $v_w = \omega/k=d\omega/dk$ is the wave speed and $v_\thh=\left(\int w_x^2\hat{f}d^3v\right)^{1/2}=\sqrt{T_{xx}^{\HO}}$ is the thermal speed. 
We note that the primitive closure yields physical wave speeds corresponding to forward and backward moving acoustic waves for specific heat ratio $\gamma=1$. However, the conservative closure yields a compound
wave speed that is physically inconsistent. The conservative stress term is physically consistent 
only in the subsonic limit, $u\ll v_\thh$. We will find that the choice of closure (i.e., conservative or primitive) for the LO system has a large effect on the performance of the overall HOLO system. A full derivation of the dispersion relations in this section is presented in \ref{sec:AppDisperion}.

\subsubsection{5-moment system}

The natural extension to the 4-moment system we can study is the 5-moment system, wherein we include the evolution equation for the stress tensor element $S_{xx}$. We may write the conservative and primitive forms of the 5-moment set of equations as 
\begin{equation}\label{eq:5MCP}
\textrm{Conservative: }
\begin{cases}
    \mathlarger{\partial_tn_{s}+\partial_x\Gamma_{s,x} = 0,} \quad \\
    \\
    \mathlarger{\partial_t\Gamma_{s,x}+\partial_xS_{s,xx}= a_{s,x},}   \\
    \\
    \mathlarger{\partial_t\Gamma_{s,\perp}+\partial_x\left(n_s \tilde{S}^\HO_{s,x\perp}\right) = a_{s,\perp},}\\
    \\
    \mathlarger{\partial_tS_{s,xx}+\partial_x\left(n_s \tilde{Q}^\HO_{s,xxx}\right) = \dot{e}_{s,xx},}
\end{cases}
\textrm{Primitive: }
    \begin{cases}
    \mathlarger{\partial_tn_{s}+\partial_x\Gamma_{s,x} = 0,} \quad \\
    \\
    \mathlarger{\partial_t\Gamma_{s,x}+\partial_xS_{s,xx} = a_{s,x},}\\
    \\
    \mathlarger{\partial_t\Gamma_{s,\perp}+\partial_x\left[n_s T^\HO_{s,x\perp}+\frac{\Gamma_{s,x}\Gamma_{s,\perp}}{n_s}\right] = a_{s,\perp},}\\
    \\
    \mathlarger{\partial_tS_{s,xx}+\partial_x\Bigg[n_s\tilde{q}^\HO_{s,xxx}+3\frac{\Gamma_{s,x} S_{s,xx}}{n_s}}\\ 
    \qquad\qquad\qquad\qquad\qquad\mathlarger{-2\frac{\left(\Gamma_{s,x}\right)^3}{\left(n_s\right)^2}\Bigg] = \dot{e}_{s,xx},}
\end{cases}
\end{equation}
where $\tilde{Q}^{\textrm{HO}}_{xxi} = \int v_x^2v_i\hat{f}d^3v$ is the density-normalized third moment, $$\tilde{q}^{\textrm{HO}}_{xxi} = \tilde{Q}^{\textrm{HO}}_{xxi}-\frac{2\Gamma_x^\HO S_{ix}^\HO+\Gamma_i^\HO S_{xx}^\HO}{\left(n^\HO\right)^2}+\frac{2\Gamma_i^\HO\left(\Gamma_x^\HO\right)^2}{\left(n^\HO\right)^3}$$ is the density-normalized third central moment (cf. heat flux), and $$\dot{e}_{s,ij} = \frac{q_s}{m_s}\left(\Gamma_{s,i} E_j+\Gamma_{s,j} E_i+\varepsilon_{ipq}S_{s,jp} B_q+\varepsilon_{jpq}S_{s,ip} B_q\right)$$ is the tensor energy deposition rate. Note that the first two terms of $\dot{e}_{ij}$ can be thought of as generalized electric power input, while the second and third terms consist of cyclotron rotation of the stress tensor in velocity space. In this case, the linear neutral fluid dispersion relation, to first order in $\tilde{q}_{xxx}$ (i.e. for small $\tilde{q}_{xxx}$) shows a more stark difference between the conservative and primitive systems:
\begin{equation}\label{eq:5M_Disp}
\begin{split}
    \textrm{Conservative: }v_w&= \left\{0,\left(\tilde{Q}_{xxx}\right)^{1/3}\right\} = \left\{0, \left[\tilde{q}_{xxx}+u_x\left(u_x^2+3v_\thh^2\right)\right]^{1/3}\right\},  \\
    \textrm{Primitive: }v_w &= \left\{u_x,u_x-\frac{\tilde{q}_{xxx}}{3v_\thh^2},u_x\pm \sqrt{3}v_\thh+\frac{\tilde{q}_{xxx}}{6v_\thh^2}\right\},
\end{split}
\end{equation}
where the cube root encompasses all three complex solutions. The conservative system results in wave speeds that are not physically consistent in any regime and includes two complex wave speeds. This is in contrast with the primitive system, which yields the physically consistent wave speeds for $\gamma=3$ in the limit of the heat flux $\tilde{q}_{xxx}\rightarrow 0$. 

\subsubsection{7-moment system}
The 7-moment system comprises the full set of second velocity moment equations required to close the first-order moment equations:
\begin{equation}\label{eq:7MC}
\begin{split}
\textrm{Conservative: }&
\begin{cases}
    \mathlarger{\partial_tn_{s}+\partial_x\Gamma_{s,x} = 0,} \quad \\
    \\
    \mathlarger{\partial_t\Gamma_{s,i}+\partial_xS_{s,ix} = a_{s,i},}\\
   \\
    \mathlarger{\partial_tS_{s,ix}+\partial_x\left(n_s \tilde{Q}^\HO_{s,ixx}\right) = \dot{e}_{s,ix},}
\end{cases}
\\
\textrm{Primitive: }&
    \begin{cases}
    \mathlarger{\partial_tn_{s}+\partial_x\Gamma_{s,x} = 0,} \quad \\
    \\
    \mathlarger{\partial_t\Gamma_{s,i}+\partial_xS_{s,ix} = a_{s,i},}\\
    \\
    \mathlarger{\partial_tS_{s,ix}+\partial_x\left[n_s \tilde{q}^\HO_{s,ixx}+\frac{2\Gamma_{s,x}S_{s,ix}+\Gamma_{s,i}S_{s,xx}}{n_s}-2\frac{\Gamma_{s,i}\left(\Gamma_{s,x}\right)^2}{\left(n_s\right)^2}\right] = \dot{e}_{s,ix},}
\end{cases}
\end{split}
\end{equation}
The waves obtained from the neutral fluid dispersion relation for the 7-moment systems are as follows:
\begin{equation}\label{eq:7M_Disp}
\begin{split}
    \textrm{Conservative: }v_w&=\left\{0,\left(\tilde{Q}_{xxx}\right)^{1/3}\right\}=\left\{0,\left[\tilde{q}_{xxx}+u_x\left(u_x^2+3v_\thh^2\right)\right]^{1/3}\right\}\\
    \textrm{Primitive: }v_w&=\left\{u_x-\frac{\tilde{q}_{xxx}}{3v_\thh^2},u_x\pm v_\thh,u_x\pm \sqrt{3}v_\thh+\frac{\tilde{q}_{xxx}}{6v_\thh^2}\right\}.
\end{split}
\end{equation}
We note a similar result to the 5-moment system, but that the primitive system now includes waves with $\gamma=1$ and $\gamma=3$ in the limit of $\tilde{q}_{xxx}\rightarrow0$. As noted in other fluid work~\cite{brown1995numerical,kuldinow2024ten}, the new ($\gamma=1$) waves are associated with transverse waves, i.e., shear transport.

\section{Discretizations and Solvers}\label{sec:Discretization}
\vspace{10pt}

In this section, we provide a detailed description of the numerical implementation of the HOLO multiscale solver.

\subsection{HO System}\label{sec:HO_discrete}
We utilize the 1-D electromagnetic particle push scheme presented in Ref.~\citenum{chen2014energy}, wherein for each particle $p$, the full timestep $\Delta t$, i.e., from timestep $\eta $ to $\eta+1$,  is subdivided into multiple substeps $\Delta \tau_p$ such that over all substeps $\nu$, $t^{\eta+1}-t^\eta = \Delta t = \sum_\nu\Delta\tau^{\nu}_p$. Omitting the particle subscript $p$ and timestep superscript $\eta$, the substep update is discretized using an implicit Crank-Nicolson scheme:
\begin{equation}\label{eq:particlepush}
        \frac{x^{\nu+1}-x^{\nu}}{\Delta\tau^\nu}=v_x^{\nu+\hlf},\qquad 
        \frac{v_i^{\nu+1}-v_i^{\nu}}{\Delta\tau^\nu}=\frac{q}{m}\left(E_i^{\eta+\hlf}+\varepsilon_{ijk}v_j^{\nu+\hlf}B_k^{\nu+\hlf}\right),
\end{equation}
where the electromagnetic fields scattered to the particle position are calculated using the methods in Ref.~\citenum{chen2014energy} and using the Crank-Nicolson discretization, $Q^{\nu+\hlf} = \left(Q^\nu+Q^{\nu+1}\right)/2$ for quantities $Q = \left\{x,v_i,B_i\right\}$. It is to be noted that the electric field scattered to the particles is at time $t^{\eta+\hlf}$ for all substeps while, to enforce conservation of particle canonical momentum, the magnetic field is calculated at time $t^{\nu+\hlf}$. The specific value of $\Delta \tau^\nu$ is chosen (i)~to minimize local error associated with stepping over the field timescales: $\Delta\tau^\nu \leq \tau_{\textrm{field}} \equiv 0.1 \times \textrm{min} \left( \omega_\T^{-1}, \omega_{\cc}^{-1}\right)$, where $\omega_\T = \sqrt{\left|q\partial_xE_x / m\right|}$ and $\omega_{\cc} = \left|qB/m\right|$~\cite{chen2014energy,chen2011energy}, and (ii)~so that the particle does not cross between cells within a substep: $x_{\ell-\hlf}-x^{\nu}\leq v_x^{\nu+\hlf}\Delta\tau^\nu\leq x_{\ell+\hlf}-x^{\nu}$, where $x_{\ell\pm\hlf}$ are the $x$-positions of the left and right interfaces of the cell, $\ell$, that the particle is in. The second condition is to ensure exact charge conservation, which can be violated if the particle traverses between cells within a substep, thereby moving outside the range of support of the shape functions used to determine $E_i\left(x_p\right)$. 

The update in Eqn.~\eqref{eq:particlepush} is performed implicitly. We employ an under-relaxed Picard iteration, advancing from the $\left(k\right)^{\rm th}$ to the $(k+1)^{\rm th}$ iteration as:
\begin{equation}\label{eq:Picard}
    \begin{split}
        x^{\nu+1,(k+1)} &= x^{\nu} + \Delta\tau^{\nu,(k)} v_x^{\nu+\hlf,(k)}, \\
        v_i^{\nu+1,(k+1)} &= v_i^{\nu} + \Delta\tau^{\nu,(k)} \frac{q}{m}
        \left[
            E_i^{\eta+\hlf,(k+1)} + \varepsilon_{ijk} v_j^{\nu+\hlf,(k)} B_k^{\nu+\hlf,(k)}
        \right].
    \end{split}
\end{equation}
The substep size for the next estimate is updated with under-relaxation:
\begin{equation}\label{eq:dtau2}
    \Delta\tau^{\nu,(k+1)} = (1 - \alpha)\Delta\tau^{\nu,(k)} + \alpha \Delta\tau^{\nu,(k+1)'},
\end{equation}
where $\alpha$ is a mixing parameter that controls the degree of under-relaxation. The estimate $\Delta\tau^{\nu,(k+1)'}$ is computed as:
\begin{equation}\label{eq:dtau1}
    \Delta\tau^{\nu,(k+1)'} = \min\left\{
        \Delta t - \sum_{\sigma=1}^{\nu-1} \Delta\tau^\sigma,\;
        \tau_{\textrm{field}},\;
        \frac{x_{\ell\pm\hlf} - x^{\nu,(k)}}{v_x^{\nu+\hlf,(k)}}
    \right\},
\end{equation}
which selects the minimum of the remaining global timestep, the field-limited timescale, and the time to reach the cell interface. The substep size computed in Eqn.~\eqref{eq:dtau2} is then used in the next iteration of Eqn.~\eqref{eq:Picard}. In practice, the Picard iteration converges within 3--6 iterations, though up to 10 may be required for large $\Delta t$ (see solver statistics in \S\ref{sec:Tests}).\par

Under-relaxation of the substep size is essential for robust iteration. Consider a particle climbing a potential barrier near a cell interface: it may either reach the interface and be forced to stop (cell-size limited), or turn around and descend the barrier, in which case it can take a larger step (field-limited). The discrete switching between these two regimes can lead to oscillations in the estimated substep $\Delta\tau$ during Picard iterations. Although such cases are rare and typically involve only a single particle, their impact can be significant, especially when many particles and timesteps are involved. A single unconverged particle can introduce a relative energy error of $1/({\rm \#\,particles})$, which may exceed the simulation’s tolerance. We found that applying a modest amount of under-relaxation ($\alpha = 0.95$) effectively suppresses these oscillations and prevents iteration stalling. This smoothing enables robust convergence of the Picard iteration, even in cases where particles are forced to stop precisely at cell interfaces.\par

Once all particles are updated through the full $\Delta t$, the moments are calculated for use in the LO system. The moments are calculated on a staggered grid, based on the order of the $v_x$-moment. The moment that are even-order in $v_x$, e.g. $n, \Gamma_y,S_{xx},$ and $Q_{xxy}$, are located at cell centers and use a second-order shape function, $S_2(x)$. For instance, the number density is calculated as
\begin{equation}\label{eq:ncalc}
    n^{\eta,\HO}_{\ell} = \sum_pw_pS_2\left(x_p^\eta-x_\ell\right),
\end{equation}
where $x_\ell$ is the $x$ position of cell center $\ell$. The half-timestep momentum densities are treated slightly differently due to their intimate connection to the fields and to ensure exact charge conservation. The moments that are odd-order in $v_x$, e.g. $\Gamma_x,S_{xy},$ and $Q_{xxx}$, are located at cell interfaces and use a first-order shape function, $S_1(x)$.  The momentum densities are \textit{orbit averaged}, as described in Refs.~\citenum{chen2011energy} and \citenum{chen2014energy}:
\begin{equation}\label{eq:GxOA}
    \Gamma_{x,\ell+\hlf}^{\eta+\hlf,\HO} = \frac{1}{\Delta t}\sum_{p,\nu}w_p\Delta\tau^{\nu}v_{p,x}^{\nu+\hlf}S_1\left(x^{\nu+\hlf}_p-x_{\ell+\hlf}\right),
\end{equation}
\begin{equation}\label{eq:GyOA}
    \Gamma_{y,\ell}^{\eta+\hlf,\HO} = \frac{1}{\Delta t}\sum_{p,\nu}w_p\Delta\tau^{\nu}v_{p,y}^{\nu+\hlf}S_2\left(x^{\nu+\hlf}_p-x_{\ell}\right),
\end{equation}
and likewise for $\Gamma_{z,\ell}^{\eta+\hlf}$. However, all moments aside from $\Gamma_i$ are evaluated at integer timesteps. The precise forms of the shape functions and how other moments are calculated are presented in \ref{sec:AppMoments}. As in previous work \cite{chen2011energy,chen2014energy}, for periodic domains, the moments and fields are binomially smoothed:
\begin{equation}\label{eq:smoothing}
    \textrm{SM}\left(\mathcal{M}\right)_\ell = \frac{\mathcal{M}^{\HO}_{\ell-1}+2\mathcal{M}^{\HO}_{\ell}+\mathcal{M}^{\HO}_{\ell+1}}{4},
\end{equation}
to reduce noise associated with high wavenumber modes introduced with particle noise and the interpolation. It has been shown in Ref.~\citenum{chen2014energy} that this smoothing does not affect the conservation properties of the simulation and allows for more robust convergence of the HOLO iteration. The computational outline of the full HO system is presented in Algorithm~\ref{alg:HO}.\par
\begin{algorithm}
\caption{HO Algorithm}\label{alg:HO}
    \begin{algorithmic}
        \Function{HighOrder}{$x_p^{\eta},v^{\eta}_{p,i},\boldsymbol{\mathcal{F}}^{\eta},\boldsymbol{\mathcal{F}}^{\eta+1}$}\Comment{Push particles based on field estimates}
        \For{$p\gets 1\ldots N_p$}
        \State $\nu \gets 0$\Comment{Initialize substep count}
        \While{$\sum_\nu\Delta\tau^\nu<\Delta t$}
        \State $k \gets 0$\Comment{Initialize Picard iteration count}
        \While{$\left|\left\{x^{(k)}-x^{(k-1)},v_i^{(k)}-v_i^{(k-1)}\right\}\right|>\textrm{tol}_{\HO}|\{\Delta x,v_{\thh}\}|$}
        \State Calculate substep size $\Delta \tau^{\nu,(k)}$ according to Eqns.~\eqref{eq:dtau2} and \eqref{eq:dtau1} 
        \State Calculate $x^{\nu+1, (k+1)},v_i^{\nu+1, (k+1)}$ using $\boldsymbol{\mathcal{F}}^{\eta}$ and $\boldsymbol{\mathcal{F}}^{\eta+1}$ according to Eqn.~\eqref{eq:Picard}
        \State $k\gets k+1$
        \EndWhile
        \State Deposit orbit-averaged $\Gamma^{\nu+\hlf,\HO}_i$ according to Eqns.~\eqref{eq:GxOA} and \eqref{eq:GyOA}
        \State $\nu\gets \nu+1$
        \EndWhile
        \EndFor
            \State \textbf{return} $\left(x_p^{\eta+1},v^{\eta+1}_{p,i}, \Gamma^{\eta+\hlf,\HO}_i\right)$
        \EndFunction
    \end{algorithmic}
\end{algorithm}

\subsection{LO System}\label{sec:LO_discrete}

\paragraph{\textbf{Discretization}} The fluid moment equations for the even-order moments are discretized using the Crank-Nicolson method to obtain the solution at time $t^{\eta+1}$. Because the field equations require the solution for $\Gamma_i^{\eta+\hlf}$, we use a half-timestep backward Euler method for the momentum evolution equations (i.e., first-order moment). For instance, the conservative 4-moment equations,  i.e., Eqn.~\eqref{eq:4MCP}, with the Darwin system, i.e., Eqns.~\eqref{eq:dtEx}-\eqref{eq:Eperp}, result in the following discretized equations for the LO system (dropping the species index):
\begin{equation}\label{eq:4M_Disc1}
    \frac{n^{\eta+1}_\ell-n^{\eta}_\ell}{\Delta t}+\frac{\Gamma_{x,\ell+\hlf}^{\eta+\hlf}-\Gamma_{x,\ell-\hlf}^{\eta+\hlf}}{\Delta x}=0,
\end{equation}
\begin{multline}\label{eq:4M_Disc2}
    \frac{\Gamma_{x,\ell+\hlf}^{\eta+\hlf}-\Gamma_{x,\ell+\hlf}^{\eta}}{\Delta t/2}+\frac{n_{\ell+1}^{\eta+\hlf}\tilde{S}_{xx,\ell+1}^{\eta+\hlf,\HO}-n_{\ell}^{\eta+\hlf}\tilde{S}_{xx,\ell}^{\eta+\hlf,\HO}}{\Delta x}\\
    -\frac{q}{m}\left(
    n_{\ell+\hlf}^{\eta+\hlf}E_{x,\ell+\hlf}^{\eta+\hlf}+\Gamma_{y,\ell+\hlf}^{\eta+\hlf}B_{z,\ell+\hlf}^{\eta+\hlf}-\Gamma_{z,\ell+\hlf}^{\eta+\hlf}B_{y,\ell+\hlf}^{\eta+\hlf}
    \right)
    -\gamma^{\eta+\hlf}_{\Gamma_x,\ell+\hlf}=0,
\end{multline}
\begin{multline}\label{eq:4M_Disc3}
    \frac{\Gamma_{y,\ell}^{\eta+\hlf}-\Gamma_{y,\ell}^{\eta}}{\Delta t/2}+\frac{n_{\ell+\hlf}^{\eta+\hlf}\tilde{S}_{xy,\ell+\hlf}^{\eta+\hlf,\HO}-n_{\ell-\hlf}^{\eta+\hlf}\tilde{S}_{xy,\ell-\hlf}^{\eta+\hlf,\HO}}{\Delta x}\\
    -\frac{q}{m}\left(
    n_{\ell}^{\eta+\hlf}E_{y,\ell}^{\eta+\hlf}+\Gamma_{z,\ell}^{\eta+\hlf}B_{x,\ell}^{\eta+\hlf}-\Gamma_{x,\ell}^{\eta+\hlf}B_{z,\ell}^{\eta+\hlf}
    \right)
    -\gamma^{\eta+\hlf}_{\Gamma_y,\ell}=0,
\end{multline}
\begin{equation}\label{eq:4M_Disc4}
    \epsilon_0\frac{E_{x,\ell+\hlf}^{\eta+1}-E_{x,\ell+\hlf}^{\eta}}{\Delta t}+
    \left(\sum_sq_s\Gamma_{s,x,\ell+\hlf}^{\eta+\hlf}-\left\langle\sum_sq_s\Gamma_{s,x,\ell+\hlf}^{\eta+\hlf}\right\rangle\right)=0,
\end{equation}
\begin{equation}\label{eq:4M_Disc5}
    \frac{A_{y,\ell+1}^{\eta+\hlf}-2A_{y,\ell}^{\eta+\hlf}+A_{y,\ell-1}^{\eta+\hlf}}{\Delta x^2}+\mu_0\left(\sum_sq_s\Gamma_{s,y,\ell}^{\eta+\hlf}-\left\langle\sum_s q_s\Gamma_{s,y,\ell}^{\eta+\hlf}\right\rangle\right)=0,
\end{equation}
and likewise for $\Gamma_{z}$ and $A_z$, where in the discrete case $\langle\cdot\rangle = \frac{1}{N_x}\sum_\ell\left(\cdot\right),$ where $N_x$ is the number of cells. In the above equations, $\gamma_{\mathcal{M}}$ is a HOLO consistency term for equation for moment ${\cal M}$. Upon convergence of the HOLO iteration, we expect all HO and LO moments to be in agreement, e.g. ${\mathcal{M}}_\ell^{\eta+1} = {\mathcal{M}}_\ell^{\eta+1,\HO} + {\cal O}\left(\epsilon_{\HO\LO}\right)$, where $\boldsymbol{\mathcal{M}}=\left\{n,\Gamma_x, \Gamma_y, \Gamma_z\right\}$ for the 4-moment system, $\boldsymbol{\mathcal{M}}=\left\{n,\Gamma_x, \Gamma_y, \Gamma_z, S_{xx}\right\}$ for the 5-moment system, $\boldsymbol{\mathcal{M}}=\left\{n,\Gamma_x, \Gamma_y, \Gamma_z,  S_{xx},  S_{xy},  S_{xz}\right\}$ for the 7-moment system, and $\epsilon_{\HO\LO}$ is the convergence tolerance for the HOLO algorithm. As an example, for $\mathcal{M}=\Gamma_x$ in Eqn.~\eqref{eq:4M_Disc2}, $\gamma_{\Gamma_x}$ is calculated as
\begin{multline}\label{eq:gammaGx}
    \gamma^{\eta+\hlf}_{\Gamma_x,\ell+\hlf}=\frac{\Gamma_{x,\ell+\hlf}^{\eta+\hlf,\HO}-\Gamma_{x,\ell+\hlf}^{\eta,\HO}}{\Delta t/2}+\frac{n_{\ell+1}^{\eta+\hlf,\HO}\tilde{S}_{xx,\ell+1}^{\eta+\hlf,\HO}-n_{\ell}^{\eta+\hlf,\HO}\tilde{S}_{xx,\ell}^{\eta+\hlf,\HO}}{\Delta x}\\
    -\frac{q}{m}\left(
    n_{\ell+\hlf}^{\eta+\hlf,\HO}E_{x,\ell+\hlf}^{\eta+\hlf}+\Gamma_{y,\ell+\hlf}^{\eta+\hlf,\HO}B_{z,\ell+\hlf}^{\eta+\hlf}-\Gamma_{z,\ell+\hlf}^{\eta+\hlf,\HO}B_{y,\ell+\hlf}^{\eta+\hlf}
    \right).
\end{multline}
It is to be noted that there are some instances in which the values of a certain moment are required at a position where they are not naturally defined. For instance, in Eqn.~\eqref{eq:4M_Disc2}, we require a value for $\Gamma_{y}$ at the cell interfaces, though it is gathered at the cell centers. For terms of this sort, we use linear reconstruction wherein, for instance, $\Gamma_{y,\ell+\hlf} = \left(\Gamma_{y,\ell}+\Gamma_{y,\ell+1}\right)/2$. Likewise, to define quantities like the number density at half-timesteps, we write $n_{\ell}^{\eta+\hlf} = \left(n_{\ell}^{\eta+1}+n_{\ell}^{\eta}\right)/2$. For brevity in the main text, the full discretized forms of the other moment systems are presented in \ref{sec:AppLO}. \par

\paragraph{\textbf{Solver}} The discretized moment and field equations, e.g., Eqns.~\eqref{eq:4M_Disc1}-\eqref{eq:4M_Disc5}, are of the form
\begin{equation}\label{eq:F=0}    
    \mathbf{F}\left(\mathbf{U}\right) = {\bf 0},
\end{equation}
where $\mathbf{F}$ is the vector objective function, $\mathbf{U} = \left\{ \mathbf{U}_1, \cdots, \mathbf{U}_{N_x} \right\}$ is the solution vector, ${\bf U}_{\ell} = \left\{ \allowbreak \boldsymbol{\mathcal{M}}_{\ell},\boldsymbol{\mathcal{F}}_{\ell}\right\}$, $\boldsymbol{\mathcal{M}}_{\ell} = \left\{n_{s_1,\ell}^{\eta+1},n_{s_2,\ell}^{\eta+1},\cdots,\Gamma_{s_1,i,\ell+1/2}^{\eta+\hlf},\cdots\right\}$ for species $s_1,s_2,\cdots$, and $\boldsymbol{\mathcal{F}}_\ell = \left\{E_{x,\ell+\hlf}^{\eta+1},A_{y,\ell}^{\eta+\hlf},A_{z,\ell}^{\eta+\hlf}\right\}$. Equation~\eqref{eq:F=0} is solved using a preconditioned Jacobian-Free Newton Krylov (JFNK) method~\cite{kelley1995iterative,knoll2004jacobian}. Supposing at Newton iteration $\lambda$,
\begin{equation}\label{eq:Residual}
    \mathbf{R}^{(\lambda)} = \mathbf{F}\left(\mathbf{U}^{(\lambda)}\right),
\end{equation}
where $\mathbf{R}$ is the vector of nonlinear residuals, which we would like to minimize through iteration. For each iteration, the solution vector is updated  as
\begin{equation}\label{eq:U_update}
    \mathbf{U}^{(\lambda+1)}=\mathbf{U}^{(\lambda)}+\mathbf{\delta U}^{(\lambda)},
\end{equation}
where $\mathbf{\delta U}$ is the Newton update, obtained by solving the following linear system: 
\begin{equation}\label{eq:FU-R}
    \mathbb{J}^{(\lambda)}\mathbf{\delta U}^{(\lambda)}=-
    \mathbf{R}^{(\lambda)}.    
\end{equation}
Here, $\mathbb{J}^{(\lambda)} = \left(\frac{\partial \mathbf{F}}{\partial\mathbf{U}}\right)^{(\lambda)}$ is the Jacobian matrix and we use GMRES to solve for $\delta {\bf U}^{(\lambda)}$. We impose a tolerance of ${\rm tol}_{\rm GMRES}=10^{-10}$ relative to the initial residual vector. We allow for an order-$100$ Krylov subspace, that is, we allow a maximum of $100$ iterations before restarting the method, and also at most $100$ outer iterations (restarts) for the method to converge.

\paragraph{\textbf{Preconditioner}} To accelerate GMRES convergence, we employ a left-preconditioning strategy. We consider a Quasi-Newton update by simplifying the electron 4-moment equations, Eqns.~\eqref{eq:4M_Disc1}-\eqref{eq:4M_Disc5}, neglecting magnetic-field Lorentz force contributions to decouple momenta from magnetic fields. This approximation yields the following Quasi-Newton system for the electron-field equations:
\begin{equation}\label{eq:4M_Pre1}
    -R_{n_\e,\ell} = \frac{\delta n_{\e,\ell}}{\Delta t}+\frac{\delta\Gamma_{\e,x,\ell+\hlf}-\delta\Gamma_{\e,x,\ell-\hlf}}{\Delta x},
\end{equation}
\begin{multline}\label{eq:4M_Pre2}
    -R_{\Gamma_{\e,x},\ell+\hlf}=\frac{\delta\Gamma_{\e,x,\ell+\hlf}}{\Delta t/2}+\frac{\delta n_{\e,\ell+1}\tilde{S}_{\e,xx,\ell+1}^{\eta+\hlf,\HO}-\delta n_{\e,\ell}\tilde{S}_{\e,xx,\ell}^{\eta+\hlf,\HO}}{2\Delta x}\\
    -\frac{q_\e}{m_\e}\left(n_{\e,\ell+\hlf}^{\eta+\hlf}\frac{\delta E_{x,\ell+\hlf}}{2}+\frac{\delta n_{\e,\ell+\hlf}}{2} E_{x,\ell+\hlf}^{\eta+\hlf}\right),
\end{multline}
\begin{multline}\label{eq:4M_Pre3}
    -R_{\Gamma_{\e,y},\ell}=\frac{\delta\Gamma_{\e,y,\ell}}{\Delta t/2}+\frac{\delta n_{\e,\ell+\hlf}\tilde{S}_{\e,xy,\ell+\hlf}^{\eta+\hlf,\HO}-\delta n_{\e,\ell-\hlf}\tilde{S}_{\e,xy,\ell-\hlf}^{\eta+\hlf,\HO}}{2\Delta x} 
    -\frac{q_\e}{m_\e}\left(n_{\e,\ell}^{\eta+\hlf}\frac{\delta A_{y,\ell}}{2}+\frac{\delta n_{\e,\ell}}{2}E_{y,\ell}^{\eta+\hlf}\right),
\end{multline}
\begin{equation}\label{eq:4M_Pre4}
    -R_{E_x,\ell+\hlf}=\epsilon_0\frac{\delta E_{x,\ell+\hlf}}{\Delta t}+ q_\e\delta\Gamma_{\e,x,\ell+\hlf},
\end{equation}
\begin{equation}\label{eq:4M_Pre5}
    -R_{A_y,\ell} = \frac{\delta A_{y,\ell+1}-2\delta A_{y,\ell}+\delta A_{y,\ell-1}}{\Delta x^2}+\mu_0 q_\e\delta\Gamma_{\e,y,\ell}.
\end{equation}\par
We note that electron timescales dominate current evolution, allowing us to neglect ion contributions in Eqns.~\eqref{eq:4M_Pre4}-\eqref{eq:4M_Pre5}. Ion equations are preconditioned by ignoring transport and source terms, e.g., $-R_{n_\ii,\ell}=\delta n_{\ii,\ell}/\Delta t$. Further, for exact inversion using a tridiagonal solver, we employ the conservative form of the closure within the preconditioner, even if the residual is computed using the primitive closure; we demonstrate that the LO solver performs efficiently despite this approximation in the preconditioner. 

We now focus on the subsystem $\{\delta n_\e,\delta \Gamma_{x,\e},\delta E_x\}$, rewriting Eqns.~\eqref{eq:4M_Pre1}, \eqref{eq:4M_Pre2}, and \eqref{eq:4M_Pre4} in operator form:
\begin{equation}\label{eq:electron_field_block}
    \begin{pmatrix}
        \mathbb{D}_{n_\e} & \mathbb{G}^+_{n_\e \Gamma_{\e,x}} & 0 \\
        \mathbb{G}^-_{\Gamma_{\e,x} n_\e} & \mathbb{D}_{\Gamma_{\e,x}} & \mathbb{D}_{\Gamma_{\e,x} E_x} \\
        0 & \mathbb{D}_{E_x \Gamma_{\e,x}} & \mathbb{D}_{E_x}
    \end{pmatrix}
    \begin{pmatrix}
        \delta n_\e \\
        \delta \Gamma_{\e,x} \\
        \delta E_x
    \end{pmatrix}
    = -
    \begin{pmatrix}
        R_{n_\e} \\
        R_{\Gamma_{\e,x}} \\
        R_{E_x}
    \end{pmatrix},
\end{equation}
where the $\mathbb{D}$ blocks are diagonal operators and $\mathbb{G}^+$ and $\mathbb{G}^-$ blocks are band matrix operators including exactly one superdiagonal and subdiagonal, respectively. Equation~\eqref{eq:electron_field_block} can be manipulated to yield a Schur complement form:
\begin{equation}\label{eq:shur_complement_4_gamma_e}
    \widetilde{\mathbb{G}}_{\Gamma_{\e,x}} \delta \Gamma_{\e,x} = -\widetilde{R}_{\Gamma_{\e,x}},
\end{equation}
with
\begin{equation}\label{eq:shur_complement_op_4_gamma_e}
    \widetilde{\mathbb{G}}_{\Gamma_{\e,x}} =
        -\mathbb{G}^-_{\Gamma_{\e,x} n_\e} \mathbb{D}^{-1}_{n_\e} \mathbb{G}^+_{n_\e \Gamma_{\e,x}}
        + \mathbb{D}_{\Gamma_{\e,x}}
        - \mathbb{D}_{\Gamma_{\e,x} E_x} \mathbb{D}^{-1}_{E_x} \mathbb{D}_{E_x\Gamma_{\e,x}},
\end{equation}
\begin{equation}\label{eq:shur_complement_res_4_gamma_e}
    \widetilde{R}_{\Gamma_{\e,x}} =
    \mathbb{G}^-_{\Gamma_{\e,x}n_\e} \mathbb{D}^{-1}_{n_\e} R_{n_\e}
    -R_{\Gamma_{\e,x}} 
    + 
    \mathbb{D}_{\Gamma_{\e,x}E_x} \mathbb{D}^{-1}_{E_x} R_{E_x}.
\end{equation}
The resulting operator $\widetilde{\mathbb{G}}_{\Gamma_{\e,x}}$ is tridiagonal, since it contains only diagonal operators and one composition of superdiagonal and subdiagonal operators; thus, it is inverted efficiently using the Thomas algorithm at ${\cal O}\left(N_x\right)$ versus larger cost if we were to directly invert Eqn. \eqref{eq:electron_field_block}. After solving for $\delta \Gamma_{\e,x}$, we substitute back into Eqns.~\eqref{eq:4M_Pre1} and \eqref{eq:4M_Pre5} to obtain $\delta n_\e$ and $\delta E_x$.\par

A similar procedure is used for the equations for $\delta\Gamma_{\e,y}$ and $\delta A_y$. Note that $\delta \Gamma_{\e,y,\ell}$ is a function of $\delta A_{y,\ell}$ only since $\delta n$ is already known. These equations are then substituted into Eqn.~\eqref{eq:4M_Pre5} to obtain another tridiagonal system for $\delta A_y$. An equivalent procedure is used to solve for $\delta A_z$ and $\delta \Gamma_{\e,z}$. In total, this requires three tridiagonal matrix solves ($\mathcal{O}(N_x)$) per GMRES iteration. Refer to Algorithm \ref{alg:LO} for a summary of the LO solver.\par

\begin{algorithm}
\caption{LO Solver Algorithm}\label{alg:LO}
    \begin{algorithmic}
        \Function{LowOrder}{$\boldsymbol{\mathcal{M}}^\eta,\boldsymbol{\mathcal{M}}^{\eta+1,\HO},\boldsymbol{\mathcal{F}}^\eta$}\Comment{Update fields based on particle moments}
        \State $\lambda\gets 0$\Comment{Initialize LO iteration count}
        \State Construct solution vector $\textbf{U}^{(\lambda=0)} = \{\boldsymbol{\mathcal{M}}^{\eta+1},\boldsymbol{\mathcal{F}}^{\eta+1}\}^{(0)}$
        \State Calculate $\textbf{R}^{(0)}$ using Eqn.~\eqref{eq:Residual}
        \While{$|\mathbf{R}^{(\lambda)}|>\epsilon_{\LO}|\mathbf{R}^{(0)}|$}
        \State Calculate $\delta\textbf{U}^{(\lambda)}$ using JFNK 
        and preconditioning explained in \S\ref{sec:LO_discrete}.
        \State $\textbf{U}^{(\lambda+1)}\gets \textbf{U}^{(\lambda)}+\delta\textbf{U}^{(\lambda)}$
        \State Calculate $\textbf{R}^{(\lambda+1)}$ using Eqn.~\eqref{eq:Residual}
        \State $\lambda\gets\lambda+1$
        \EndWhile
        \State \textbf{return} $\left\{\boldsymbol{\mathcal{F}}^{\eta+1}\right\}$
        \EndFunction
    \end{algorithmic}
\end{algorithm}

\paragraph{\textbf{5- and 7-moment systems}} The 5- and 7-moment LO systems cannot be coerced into a tridiagonal form as easily. Even neglecting magnetic couplings, the 5-moment (Eqns.~\eqref{eq:5MCP},\eqref{eq:5M_n}-\eqref{eq:5M_Sxx}) and 7-moment systems (Eqns.~\eqref{eq:7MC},\eqref{eq:7M_n}-\eqref{eq:7M_Sxz}) would require the solution of penta-diagonal matrices. Instead, for the 5- and 7-moment systems, we keep the HO closure in both the $\Gamma$ and $S$ equations. So, the exact same procedure is followed above to solve for $\{\delta n_\e, \delta\Gamma_{\e,i},\delta E,\delta A_i\}$. Then, the equation for $S_{\e,xx}$, Eqn.~\eqref{eq:5M_Sxx}, can be turned into a linear update equation as
\begin{multline}\label{eq:dSxx}
    -R_{S_{\e,xx},\ell} = \frac{\delta S_{\e,xx,\ell}}{\Delta t}
    +\frac{\delta n_{\e,\ell+\hlf}\tilde{Q}_{\e,xxx,\ell+\hlf}^{\eta+\hlf,\HO}-\delta n_{\e,\ell-\hlf}\tilde{Q}_{\e,xxx,\ell-\hlf}^{\eta+\hlf,\HO}}{\Delta x}
    -\frac{2q_\e}{m_\e}\Bigg(
    \delta\Gamma_{\e,x,\ell}E_{x,\ell}^{\eta+\hlf}+\Gamma_{\e,x,\ell}^{\eta+\hlf}\delta E_{x,\ell}
    \\+
    \delta n_{\e,\ell}S_{\e,xy,\ell}^{\eta+\hlf,\HO}B_{z,\ell}^{\eta+\hlf}+
    n_{\e,\ell}^{\eta+\hlf}S_{\e,xy,\ell}^{\eta+\hlf,\HO}\delta B_{z,\ell}
    -
    \delta n_{\e,\ell}S_{\e,xz,\ell}^{\eta+\hlf,\HO}B_{y,\ell}^{\eta+\hlf}
    -
    n_{\e,\ell}^{\eta+\hlf}S_{\e,xz,\ell}^{\eta+\hlf,\HO}\delta B_{y,\ell}
    \Bigg),
\end{multline}
where $\delta B_{i,\ell}=\epsilon_{ixj}(\delta A_{j,\ell+\hlf}-\delta A_{j,\ell-\hlf})/\Delta x$. Since the closure and other $S_{\e,ij}$ terms use the HO values, there are no unknown terms in this equation, and Eqn.~\eqref{eq:dSxx} can be trivially solved for $\delta S_{\e,xx,\ell}$. It is found that the off-diagonal stress elements do not impose stiff timescales, so they are also solved simply as $-R_{S_{\e,xy},\ell+\hlf} = \delta S_{\e,xy,\ell+\hlf}/\Delta t$, and likewise for $\delta S_{\e,xz}$. This is to be expected because the transverse shear waves travel at speeds strictly slower than longitudinal pressure waves, as noted in Eqn.~\eqref{eq:7M_Disp}. Furthermore, since the density and momenta are already solved, the only additional information provided by the off-diagonal stress tensor equations is of the off-diagonal pressures, which are close to zero. Thus, to a good approximation, the changes in $S_{xy}$ and $S_{xz}$ are fully determined by the updates of $n$ and $\Gamma_i$. The development of more sophisticated preconditioners for the LO system may be required for coupled problems with strong shear effects and even more complex moment systems, but is reserved for future work.

\subsection{Anderson Accelerated HOLO Iteration}

The HOLO iteration consists of the HO system, which solves for particle positions and velocities, from which moments are gathered and used in the LO system, which solves for the electromagnetic fields. We accelerate the coupled HOLO convergence by leveraging Anderson mixing~\cite{anderson1965iterative,walker2011anderson,willert2014leveraging}. Since the HO system is linear for a given evaluation of the fields, $\left\{E_x, A_y, A_z \right\}$, we define the HOLO residual, i.e., what we want to minimize, as the change in the HO fluid moments and LO fields obtained from one HOLO iteration to the next. That is,
\begin{equation}\label{eq:HOLO_res}
    \textbf{r}^{(y+1)} = \textbf{U}^{\HO,(y+1)}-\textbf{U}^{\HO,(y)},
\end{equation}
where $y$ is the HOLO iteration index and ${\bf U}^{\HO,(y)} = \left\{\boldsymbol{\cal M}^{\HO,(y)}, \boldsymbol{\cal F}^{(y)} \right\}$.\par
We have found that the Anderson mixing algorithm was greatly beneficial in improving robust convergence of the HOLO iteration, particularly for large time-step sizes.  We use the Anderson mixing algorithm, presented in Algorithm~\ref{alg:Anderson} in \ref{sec:AppAA} for completeness, which takes in the HOLO residuals at the last $h$ iterations, $\{\textbf{r}^{(y)},\textbf{r}^{(y-1)},\cdots,\textbf{r}^{(y-h+1)}\}$, and obtains a new estimate of the $\textbf{U}^{\HO,(y)}$. The estimates of the fields are then passed into the HOLO update. This iteration is repeated until the norm of the residual (i.e., the HOLO update) falls below a user-specified tolerance, $\epsilon_{\HO\LO}$. Once this occurs, the HO system is performed one more time to ensure exact conservation properties with the new estimates of the fields before moving on to the next timestep. The combined Anderson accelerated HOLO algorithm is presented in Algorithm~\ref{alg:HOLO}.

\begin{algorithm}
\caption{HOLO Algorithm. The \textsc{HighOrder}, \textsc{LowOrder}, and \textsc{AndersonMixing} functions are defined in Algorithm \ref{alg:HO}, \ref{alg:LO}, and \ref{alg:Anderson}, respectively. }
\label{alg:HOLO}
    \begin{algorithmic}
        \Procedure{HOLOAlgorithm}{$x_p^{\eta},v_{p,i}^{\eta},\boldsymbol{\cal F}^{\eta}$}\Comment{Given solution from previous time}
        \State $y\gets 0$\Comment{Initialize HOLO iteration index}
        \State $\boldsymbol{\mathcal{F}}^{\eta+1,(y=0)}=\boldsymbol{\mathcal{F}}^{\eta}$
        \While{$y<1$ \textbf{or} $||\textbf{r}^{(y+1)}||_\infty>\epsilon_{\textrm{HOLO}}||\textbf{r}^{(1)}||_\infty$}
        \State $\left(x_p^{\eta+1,(y+1)},v^{\eta+1,(y+1)}_{p,i}\right)\gets $\textsc{HighOrder}$\left(x_p^{\eta},v^{\eta}_{p,i},\boldsymbol{\mathcal{F}}^{\eta},\boldsymbol{\mathcal{F}}^{\eta+1,(y)}\right)$
        \State Calculate $\boldsymbol{\mathcal{M}}^{\eta+1,(y+1)}$ using, e.g., Eqns.~\eqref{eq:ncalc}, \eqref{eq:smoothing}, and \eqref{eq:gammaGx}, and those in \ref{sec:AppMoments}.
        \State $\boldsymbol{\mathcal{F}}^{\eta+1,(y+1)} \gets $\textsc{LowOrder}$\left(\boldsymbol{\mathcal{M}}^{\eta},\boldsymbol{\mathcal{M}}^{\eta+1,(y+1)},\boldsymbol{\mathcal{F}}^{\eta}\right)$
        \State $\textbf{U}^{(y+1)'} \gets \{\boldsymbol{\mathcal{M}}^{\eta+1,(y+1)},\boldsymbol{\mathcal{F}}^{\eta+1,(y+1)}\}$
        \State $\mathbf{r}^{(y+1)} \gets \textbf{U}^{(y+1)'}-\textbf{U}^{(y)}$
        \State $\textbf{U}^{(y+1)} \gets$ \textsc{AndersonMixing}($\mathbf{r},\textbf{U},y,h$)         \State $y\gets y+1$
        \EndWhile
        \State $\left(x_p^{\eta+1},v^{\eta+1}_{p,i}\right)\gets $\textsc{HighOrder}$\left(x_p^{\eta},v^{\eta}_{p,i},\boldsymbol{\mathcal{F}}^{\eta},\boldsymbol{\mathcal{F}}^{\eta+1,(y)}\right)$
        \State Calculate $\boldsymbol{\mathcal{M}}^{\eta+1}$ using, e.g., Eqns.~\eqref{eq:ncalc}, \eqref{eq:smoothing}, and \eqref{eq:gammaGx}, and those in \ref{sec:AppMoments}.
        \State Store $\textbf{U}^{\eta+1} \gets \{\boldsymbol{\mathcal{M}}^{\eta+1},\boldsymbol{\mathcal{F}}^{\eta+1}\}$
        \EndProcedure
    \end{algorithmic}
\end{algorithm}

\section{Numerical Tests}\label{sec:Tests}
\vspace{10pt}

In this section, we present the results from three classical plasma problems: electrostatic Landau damping and the electromagnetic electron and ion Weibel instabilities. The electron Landau damping test serves to benchmark our method against previous work~\cite{chen2011energy,taitano2013development} and to compare the different LO systems; while the wide range of problems allows us to demonstrate the robust performance of the HOLO solver.
For the electrostatic test cases, the particles are initialized by sampling from an isotropic, isothermal drifting Maxwellian distribution for each species, $s$:
\begin{equation}\label{eq:fES}
    f^{\textrm{ES}}_s(x,v;t=0) 
    = 
    \frac{n_s(x;t=0)}{(2\pi)^{3/2} v_{s,\textrm{th}}^3}
    \exp\left[-\frac{\left(v_i-u_{s,i}(x;t=0)\right)^2}{2v_{s,\textrm{th}}^2}\right],
\end{equation}
where the initial density and bulk velocity are assumed to be sinusoidal perturbations in a uniform plasma:
\begin{equation}
    n_s(x;t=0) = n_0+\Delta n_s\cos\left(k_xx\right),
    \hspace{20pt}
    u_{s,i}(x;t=0) = u_{0,i}+\Delta u_{s,i}\cos\left(k_xx\right),
\end{equation}
where $\Delta n_s$ and $\Delta u_{s,i}$ are the magnitudes of the perturbations, and $k_x$ is the wavenumber of the perturbation. The electrostatic results we present are normalized temporally by the electron inverse plasma frequency, $\omega_{\p\e}^{-1}$, and spatially by the electron Debye length, $\lambda_{\textrm{De}}$.
\par
For the electromagnetic test cases, particles are sampled from an anisotropic isothermal drifting Gaussian distribution for each species
\begin{equation}\label{eq:fEM}
    f^{\textrm{EM}}_s(x,v_i;t=0) = \frac{n_s(x)}{(2\pi)^{3/2}v_{s,\textrm{th},x}v_{s,\textrm{th},\perp}^2}\exp\left[-\frac{\left(v_x-u_{s,x}(x)\right)^2}{2v_{s,\textrm{th},x}^2}
    -
    \frac{\left(v_y-u_{s,y}(x)\right)^2
    +\left(v_z-u_{s,z}(x)\right)^2}{2v_{s,\textrm{th},\perp}^2}\right],
\end{equation}
where $v_{s,\textrm{th},i}$ is the thermal velocity in the $i$ direction $\left(v_{s,\textrm{th},i}^2 = \int (v_i-u_i)^2\hat{f}d^3v \equiv kT_i/m\right)$ and $v_{s,\thh,y}=v_{s,\thh,z}=v_{s,\thh,\perp}$.  The electromagnetic results we present are normalized temporally by the electron inverse plasma frequency, $\omega_{\p\e}^{-1}$, and spatially by the electron skin depth $d_\e=c/\omega_{\p\e}$.\par 

In all cases, we set an average number of particles per cell, $N_\ppc$, and calculate the actual number of particles per cell, $N_{\ppc,\ell} = N_{\ppc}\lfloor n_s(x_\ell)/\bar{n}_s+1/2\rfloor$, where $\bar{n}_s$ is the average density, and use a low-discrepancy 4-dimensional Hammersley set~\cite{hammersley2013monte} of size $N_{\ppc,\ell}$ to initialize the position and velocity, i.e., $\{x,v_i\}$, of  the particles in each cell. Unless otherwise specified, we use particle Picard tolerance, $\epsilon_\HO = 10^{-12}$, LO residual (relative) tolerance, $\epsilon_\LO = 10^{-12}$, outer HOLO tolerance, $\epsilon_{\HO\LO} = 10^{-8}$, and retain $h=5$ Anderson histories (see Algorithm~\ref{alg:Anderson}). All numbers in the computational model use double precision and simulations are run on the Sherlock HPC cluster at Stanford University.

\subsection{Electrostatic electron Landau damping}

\paragraph{\textbf{Verification}} Linear Landau damping was chosen as one of the basic test problems, being an electrostatic problem that evolves over the electron timescale. The purpose of this test case is to (i) benchmark our algorithm against in the electrostatic limit and (ii) to highlight the superiority of the {primitive} formulation over the {conservative} formulation of the LO system. Given initially Maxwellian electrons and in the limit of immobile (heavy) ions, the dispersion relation from linear kinetic theory can be given by~\cite{stix1992waves}:
\begin{equation}
    1+\frac{1}{k_x^2}\left[1+\frac{\omega}{\sqrt{2}k_x}Z\left(\frac{\omega}{\sqrt{2}k_x}\right)\right]=0,
\end{equation}
where $Z$ is the plasma dispersion relation assuming a Maxwellian VDF, which can be evaluated according to Fried and Conte~\cite{fried2015plasma}. For this test case, in comparison to Refs.~\citenum{chen2011energy} and \citenum{taitano2013development}, we use a periodic domain size of $L_x=4\pi$, $k_x = 2\pi/L_x$, have a number of cells $N_x=32$, and have an average number of particles per cell $N_{\ppc} = 2500$.\par

\begin{figure}[h!]
    \centering
    \includegraphics[width = 300pt]{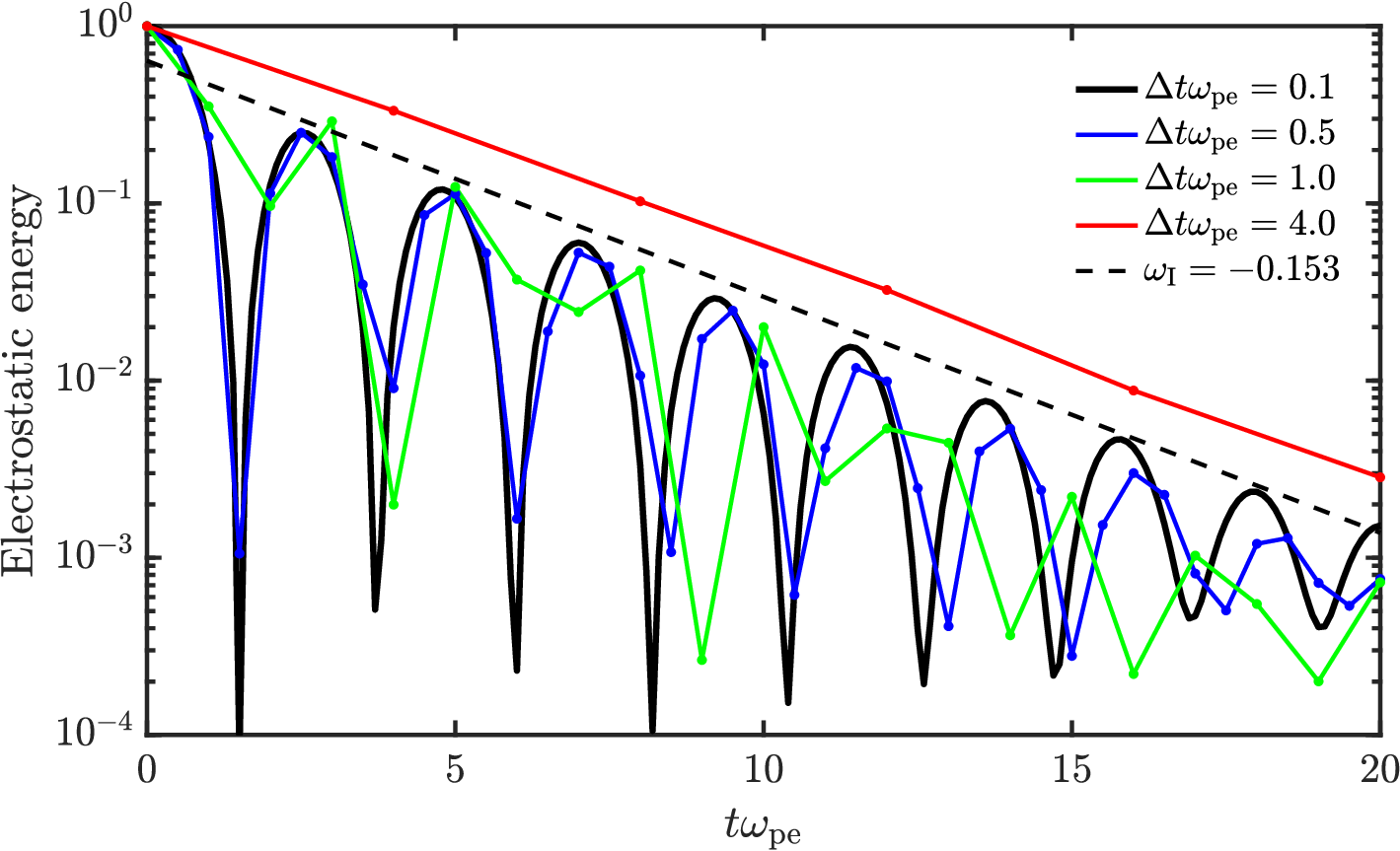}
    \caption{Electrostatic Landau damping: Plot of electrostatic energy as a function of time comparing the solutions for various timestep sizes to the analytic decay rate. These results use the 4-moment primitive formulation of the LO system and $N_{\ppc} = 2500$.}
    \label{fig:LD_NRG}
\end{figure}

The plasma is initialized with stationary warm electrons, $u_\e=0$, $T_\e = 1$ and heavy, stationary, cold ions, $u_\ii = T_\ii = 0$, $m_\ii/m_\e = 1836$. The initial density perturbation is only in the electrons, $\Delta n_{\e} = 0.01$, $\Delta n_{\ii} = 0$. For these parameters, the dispersion relation gives complex frequency $\omega = 1.416-0.153i$, giving an oscillation frequency of $\Re(\omega) = \omega_\textrm{R} = 1.416$ and a Landau damping rate of $\Im(\omega) = \omega_\textrm{I} = -0.153$. We present the results from four different choices of the timestep, $\Delta t = 0.1$, $0.5$, $ 1.0$,  and $4.0$. The plot of the electrostatic energy as a function of time is presented in Figure~\ref{fig:LD_NRG}. The electrostatic energy is calculated as $\epsilon_0\sum_{\ell = 1}^{N_x} E_{\ell-1/2}^2\Delta x/2$, and normalized by the value at time $t=0$. As demonstrated in previous work~\cite{taitano2013development}, the method is able to accurately resolve dynamical timescales up to $|\Delta t\omega|\simeq 1/2$. The oscillation frequency is still well resolved for $\Delta t\omega_\textrm{R} \simeq 0.7$, and the decay rate is still well resolved for $\Delta t\omega_\textrm{I}\simeq 0.6$, far larger than the typical explicit timestep of $\Delta t\omega_{\p\e}\leq 0.1$. The results shown use the 4-moment primitive formulation of the LO system; but, as the LO system serves as an algorithmic accelerator, the macroscopic results for all LO systems agree to within the tolerance of the HOLO iteration.\par

\paragraph{\textbf{Comparison of LO Systems}} Now, we provide a comparison of the solver performance for various choices of the LO system. Table~\ref{tab:LD_HOLO} presents the average number of HOLO iterations (itns.) per timestep, \linebreak $\left(\textrm{Total HOLO itns.}\right)/\left(\textrm{\# of  timesteps}\right)$, to reach time $t=20$ for various timestep sizes, comparing the conservative and primitive closures. The number of HOLO iterations is an important metric because it is closely tied to the number of LO iterations (field solves) and HO iterations (particle pushes), which constitute the bulk of the algorithm's computational cost. However, instead of analyzing computational runtime, which can be affected by factors such as code optimization and parallelization, each HOLO iteration includes a particle push and a global solution of the fields, and should be agnostic to implementation. Although the absolute expense varies by timestep (due to the increased subcycling required for higher $\Delta t$), for a particular value of $\Delta t$, the average number of HOLO iterations is approximately proportional to the computational expense.\par

\begin{table}[h!]
    \centering
    \caption{Electrostatic Landau damping: Average HOLO iterations per timestep for three choices of LO system, comparing conservative (left) and primitive (right) closures for various $\Delta t$ with $N_{\ppc} = 2500$.}
    \def\arraystretch{1.4}
    \begin{tabular}{|c||c|c|c|c|}
    \hline
     $\Delta t\omega_{\p\e}$&0.1&0.5&1.0&4.0  \\
     \hhline{|=||=|=|=|=|}
     4MC& 4.00 & 5.10&9.95 &8.20\\
     \hline
     5MC& 4.00 & 6.03&7.90&39.0\\
     \hline
     7MC& 4.00 & 6.15&7.90&39.0\\
     \hline
\end{tabular}
\hspace{25pt}
\begin{tabular}{|c||c|c|c|c|}
    \hline
     $\Delta t\omega_{\p\e}$&0.1&0.5&1.0&4.0  \\
     \hhline{|=||=|=|=|=|}
     4MP& 4.00& 5.15&6.15& 7.00\\
     \hline
     5MP& 4.00 & 5.15&5.30&8.60\\
     \hline
     7MP& 4.00 & 5.05&5.45&9.00\\
     \hline
\end{tabular}\label{tab:LD_HOLO}
\end{table}

For small timestep sizes, $\Delta t\omega_{\p\e} = 0.1$, the choice of LO system does not significantly impact the convergence of the coupled HO-LO system. This reflects the fact that the electric field undergoes minimal change from one timestep to the next, resulting in weak coupling to the moment system and minimal influence on the solution of the field equation. However, at larger timesteps, $\Delta t\omega_{\p\e} = 1.0$ and $\Delta t\omega_{\p\e} = 4.0$, notable differences between LO systems emerge.\par

We first observe that the conservative and primitive formulations of the 4-moment system have only a small difference in performance. This can be understood by considering the dispersion relations they admit, cf.~Eqn.~\eqref{eq:4M_Disp}. Since the bulk velocity is much smaller than the thermal speed, the two dispersion relations become nearly equivalent, leading to similar behavior in both LO systems. The primitive formulation may offer slightly improved performance due to its more physically accurate wave speeds.\par

In contrast, the 5-moment and 7-moment systems show a significantly greater disparity between conservative and primitive formulations. As seen from their respective dispersion relations, Eqns.~\eqref{eq:5M_Disp} and \eqref{eq:7M_Disp}, the conservative formulations fail to admit the correct physical wave speeds, and instead predict complex wave speeds. On the other hand, the primitive formulations accurately capture the thermal wave speed with $\gamma = 3$. This discrepancy becomes increasingly consequential as the timestep grows, since the LO system must capture a richer set of dynamics to provide an effective implicit prediction for the fields that drive the HO particle solve. For example, at $\Delta t\omega_{\p\e} = 0.5$, the use of the primitive formulation results in approximately a 15\% reduction in HOLO iterations. At $\Delta t\omega_{\p\e} = 1.0$, this reduction grows to around 30\%, and at $\Delta t\omega_{\p\e} = 4.0$, the primitive formulation requires nearly 80\% fewer HOLO iterations compared to its conservative counterpart.\par

We also find that for intermediate timestep sizes, such as $\Delta t\omega_{\p\e} = 0.5$ and $1.0$, the 5- and 7-moment systems using primitive variables outperform the 4-moment system, likely due to their ability to capture the faster $\gamma = 3$ wave mode. However, for $\Delta t\omega_{\p\e} = 4.0$, their performance degrades, possibly due to amplified numerical noise stemming from the strongly nonlinear terms in the primitive 5- and 7-moment equations, Eqns.~\eqref{eq:5MCP} and \eqref{eq:7MC} (discussed further in section \ref{subsec:e_weibel}).

\begin{table}[h!]
    \centering
    \caption{Electrostatic Landau damping: Solver statistics with the 5-moment LO system with the primitive (P) and conservative (C) closures and various $\Delta t$ for $N_{\ppc} = 2500$.}
    \def\arraystretch{1.4}
    \begin{tabular}{|c||c|c|c|c||c|c|c|c|}
    \hline
     LO|$\Delta t\omega_{\p\e}$&C|0.1&C|0.5&C|1.0&C|4.0&P|0.1&P|0.5&P|1.0&P|4.0 \\
     \hhline{|=||=|=|=|=||=|=|=|=|}
     $\left(\frac{\textrm{Picard itns.}}{\textrm{Particle substep}}\right)_{\rm avg}$& 3.0& 3.1&3.4& 4.0 &  3.0& 3.1&3.4& 4.0\\
     \hline
     $\left(\frac{\textrm{Particle substeps}}{\textrm{HOLO itn.}}\right)_{\rm avg}$& 1.5 & 3.1&5.2&17.3 & 1.5 & 3.1&5.5&18.8\\
     \hline
     $\left(\frac{\textrm{LO itns.}}{\textrm{HOLO itn.}}\right)_{\rm avg}$& 1.25 & 1.38&1.47&1.51 & 1.25 & 1.32&1.54&1.73\\
     \hline
     $\left(\frac{\textrm{HOLO itns.}}{\textrm{Timestep}}\right)_{\rm avg}$& 4.0 & 6.0&7.9&39.0 & 4.0 & 5.2&5.3&8.6\\
     \hline
     Runtime& 1 (201.0 s) & 0.491&0.479&1.608 & 0.972 & 0.417&0.328&0.386\\
     \hline
\end{tabular}\label{tab:LD_Solver}
\end{table}

\paragraph{\textbf{Solver Statistics}} The algorithmic performance of the HOLO scheme using two LO system variants (5-moment conservative and primitive forms) is shown in Table~\ref{tab:LD_Solver}. Both closure approaches exhibit nearly identical behavior, \textit{except} in the number of HOLO iterations, as previously discussed. This indicates that the HO and LO systems are individually similar in both cases.  Thus, for a given timestep size, the number of HOLO iterations is approximately proportional to the runtime.\par
The first row in Table~\ref{tab:LD_Solver} reports the average number of HO Picard iterations per particle substep, computed over all particles, substeps, and HOLO iterations. This value increases slightly with $\Delta t$ due to more frequent cell crossings, which require additional iterations to resolve interface stops. The second row shows the average number of substeps per particle per timestep. This quantity increases significantly with $\Delta t$, reflecting the need to subcycle fast particles and accommodate increased cell crossings. However, the growth is sublinear, resulting in a computational benefit from larger timesteps due to fewer total particle pushes. The third row gives the average number of LO iterations per HOLO iteration, which also increases slowly with $\Delta t$ as field updates become more significant. The fourth row presents the average number of HOLO iterations per timestep, which increases with $\Delta t$ due to stronger nonlinearities from field-particle interactions that must be converged. The final row shows the total simulation runtime, normalized to the $\Delta t\omega_{\p\e} = 0.1$ conservative case. The runtime can be approximated as:
\begin{align}\begin{split}\label{eq:Runtime}
    \textrm{Runtime} \simeq \Bigg[N_{p}&\times\frac{\textrm{CPU time}}{\textrm{Picard itn.}}\times\frac{\textrm{Picard itns.}}{\textrm{Substep}}\times\frac{\textrm{Particle substeps}}{\textrm{HOLO itn.}}\times\frac{\textrm{HOLO itns.}}{\textrm{Timestep}}\\&+\frac{\textrm{CPU time}}{\textrm{LO itn.}}\times\frac{\textrm{LO itns.}}{\textrm{HOLO itn.}}\times\frac{\textrm{HOLO itns.}}{\textrm{Timestep}} +\frac{\textrm{CPU time}}{\textrm{Incidental operations}}\Bigg] 
    \times \#\textrm{Timesteps}.
\end{split}\end{align}
Let us consider the terms in the HO contribution, i.e., the first line of Eqn.~\eqref{eq:Runtime}, one by one. Results are presented for a fixed number of particles ($N_\p$) and identical implementation (CPU time/Picard iteration) so the change in runtime in Table~\ref{tab:LD_Solver} comes mainly from the last three terms in the first line. From Table~\ref{tab:LD_Solver}, the these three terms increase with $\Delta t$, while the number of timesteps scales inversely with $\Delta t$. Consequently, the HO runtime ceases to decrease with $\Delta t$ when these three terms grow faster than $1/\Delta t$, which is highly problem-dependent. Table~\ref{tab:LD_Solver} also shows that the number of LO iterations per HOLO iteration remains nearly constant over a wide range of $\Delta t$, demonstrating the effectiveness of our preconditioning strategy. For large $\Delta t$ and $N_\p$, the HO cost dominates the total runtime. We note that incidental operations such as moment gathering, data storage, and Anderson mixing are largely independent of $\Delta t$. Further, given that the primitive closure yields superior performance in terms of HOLO iterations, lower CPU runtime, and more robust convergence at larger $\Delta t$ compared to the conservative closure, we adopt the primitive closure for the remainder of this study.

\subsection{Electromagnetic electron Weibel instability}
\label{subsec:e_weibel}

\paragraph{\textbf{Verification}}The electromagnetic electron Weibel instability is selected as a test problem characterized by dynamics on the electron timescale, with parameters chosen such that the growth rate remains small relative to $\omega_{\p\e}$. Although the instability is driven by electron-scale physics, its temporal evolution unfolds over many plasma periods, $\omega_{\p\e}^{-1}$. This makes it an ideal scenario for demonstrating the HOLO algorithm's capability to stably advance solutions with timestep sizes approaching the system’s dynamical timescale. \par

Given an initially anisotropic-Maxwellian plasma (Eqn.~\eqref{eq:fEM}), the Weibel dispersion relation from kinetic theory is~\cite{krall1973principles}:
\begin{equation}\label{eq:WeibelDisp}
    1-\frac{k_x^2c^2}{\omega}-\sum_s\frac{\omega_{\p s}^2}{\omega^2}\left[1+\frac{v_{s,\textrm{th},\perp}^2}{2v_{s,\textrm{th},x}^2}Z'\left(\frac{\omega/k_x}{\sqrt{2}v_{s,\textrm{th},x}}\right)\right]=0,
\end{equation}
where the wavenumber of the perturbation is assumed to be in the $x$ direction. 
For this test case, in comparison to Ref.~\citenum{chen2014energy}, we use a periodic domain of size $L_x = 32$, number of cells $N_x =32$, and number of particles per cell $N_{\ppc}=3000$, unless otherwise specified. The electrons are initialized as a hot, uniform, anisotropic-Gaussian plasma with $v_{\e,\textrm{th},x} = 0.025$ and $v_{\e,\textrm{th},\perp} = 0.04$, and $n_\e(x;t=0) = 1$. To introduce an initial perturbation, the electron velocity is  shifted as $u_{\e,x}(x;t=0)= 2\times10^{-5}\cos(k_xx)$, where the wavenumber of perturbation is taken to be $k = 2\pi/L_x$. For the electron Weibel instability, the ions are assumed to be hot, Maxwellian, stationary and uniform with $n_\ii(x;t=0) = 1$, $u_{\ii,i}(x;t=0) = 0$ and $v_{\ii,\textrm{th},x} = v_{\ii,\textrm{th},\perp} = 0.025$ with $m_\ii/m_\e = 1836$. For these parameters, and with ions and electrons as the only two species, the dispersion relation predicts a growth rate of $\omega_\textrm{I} = 0.004$. That is, we expect an $e$-folding in the magnitude of the perturbation every $250\textrm{ }\omega_{\p\e}^{-1}$. The simulation is run up to time $t = 2000\textrm{ }\omega_{\p\e}^{-1}$.\par 

\begin{figure}[h!]
    \centering
    \includegraphics[width = 300pt]{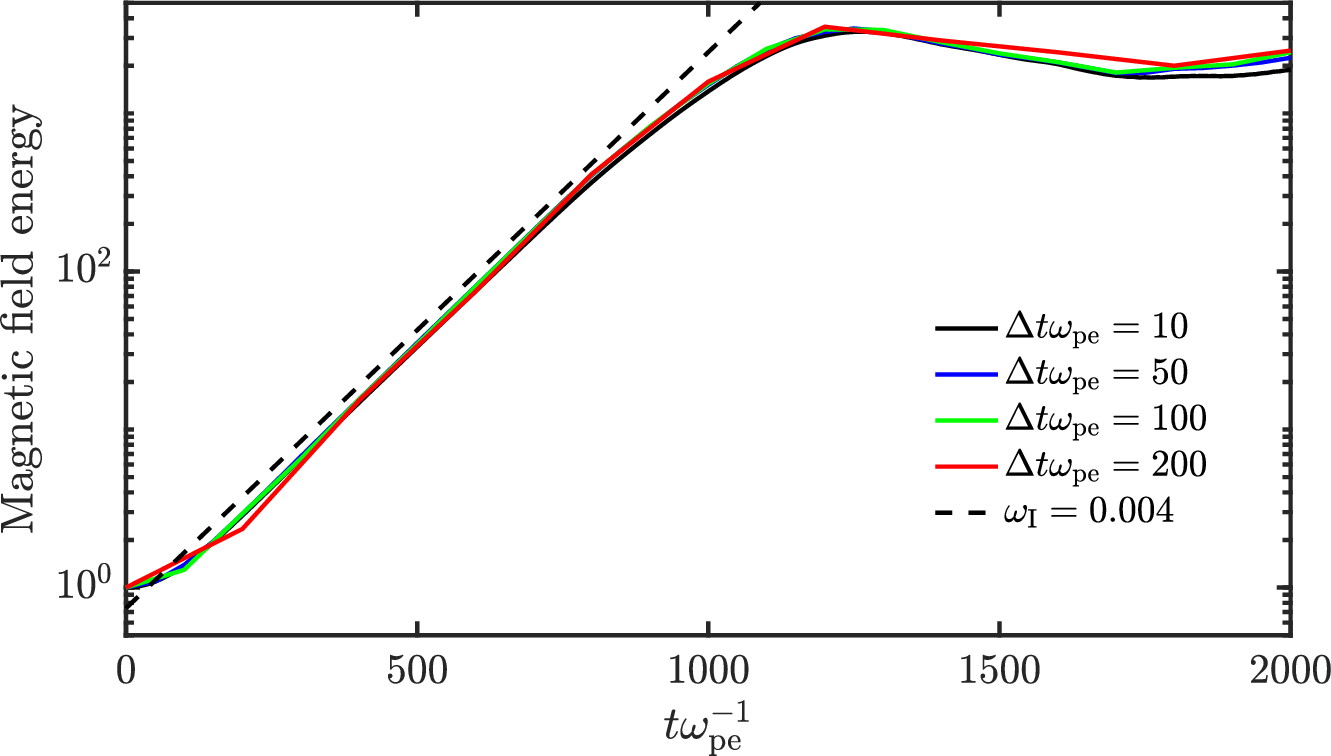}
    \caption{Electromagnetic electron Weibel instability: Plot of the magnetic field energy as a function of time, comparing the solutions for various timestep sizes to the analytic growth rate. These results use the 4-moment primitive formulation of the LO system with $N_{\ppc} = 3000$.}
    \label{fig:EW_NRG}
\end{figure}

Figure~\ref{fig:EW_NRG} presents the magnetic field energy as a function of time for various values of $\Delta t$. The magnetic energy is computed as $\sum_\ell B_\ell^2 \Delta x / 2\mu_0$ and normalized by its initial value at $t = 0$. The results show that the model accurately captures the growth rate for timestep sizes up to $\Delta t\omega_{\p\e} = 200$, or $\Delta t\omega_\mathrm{I} =0.8$, which is $2000$ times larger than the explicit stability limit of $\Delta t\omega_{\p\e} \leq 0.1$ and is just barely resolving the inverse growth rate. This highlights the HOLO solver's capability to resolve system dynamics accurately at large timesteps, without incurring numerical errors or instabilities.

\begin{figure}[h!]
    \centering
    \includegraphics[width = 280pt]{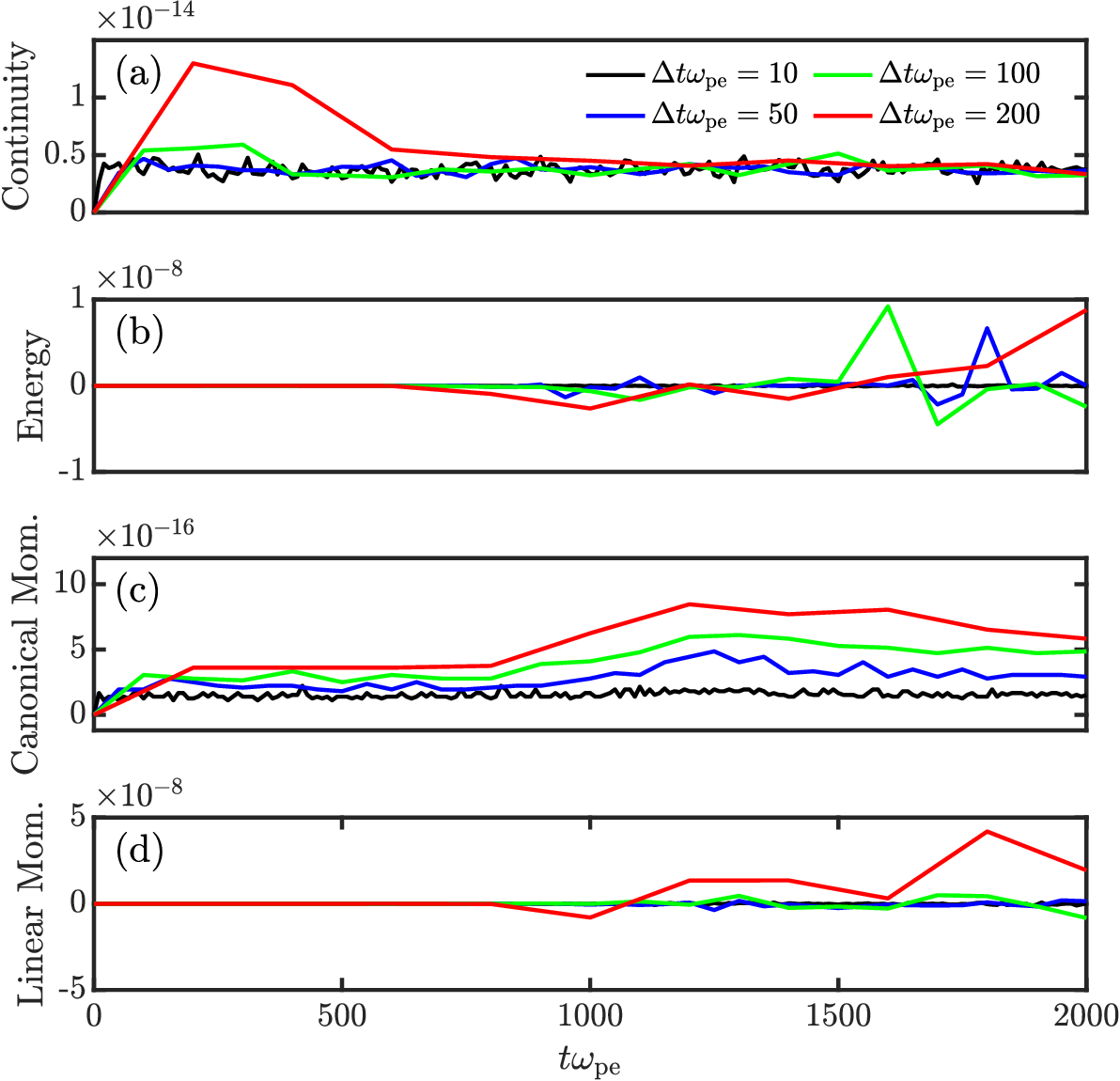}
    \caption{Electromagnetic electron Weibel instability: Conservation properties of the scheme in the electron Weibel test case for various timestep sizes. Errors are calculated as described in Eqns.~\eqref{eq:err_cont}-\eqref{eq:err_CM}}.
    \label{fig:EW_Cons}
\end{figure}

Figure~\ref{fig:EW_Cons} shows the conservation properties of the model for electromagnetic test problems, demonstrating that the timestep size does not tamper with the the conservation properties of the solver. The conservation errors for continuity, energy, (perpendicular) canonical momentum, and linear momentum are calculated as
\begin{equation}\label{eq:err_cont}
    \textrm{err}_{\textrm{Cont}} = \sqrt{\sum_{\ell}\left[\rho_\ell^{\eta+1}-\rho_\ell^\eta+\frac{\Delta t}{\Delta x}\left(j_{x,\ell+\hlf}^{\eta+\hlf}-j_{x,\ell-\hlf}^{\eta+\hlf}\right)\right]^2},
\end{equation}
\begin{equation}\label{eq:err_E}
    \textrm{err}_{\textrm{Ener}} = \frac{\mathcal{E}^{\eta+1}-\mathcal{E}^\eta}{\mathcal{E}(t=0)},
\end{equation}
\begin{equation}\label{eq:err_mom}
    \textrm{err}_{\textrm{Mom}} = \frac{\sum_p m_p v^{\eta}_{x,p}}{\sum_sm_s\bar{v}_s \left(t = 0\right)},
\end{equation}
\begin{equation}\label{eq:err_CM}
    \textrm{err}_{\textrm{CM}}=\max_p\left|m_pv_{p,\perp}^{\eta+1}+q_pA^{\eta+1}_{p,\perp}-m_pv_{p,\perp}^{\eta}+q_pA^{\eta}_{p,\perp}\right|.
\end{equation}
where the total energy is the sum of electrostatic, kinetic, and magnetic field energies:
\begin{equation}
    \mathcal{E}_\mathrm{EM} = \mathcal{E}_{E}+\mathcal{E}_{B}+\mathcal{E}_\mathrm{K} = \frac{\epsilon_0}{2}\sum_\ell\Delta x E_{x,\ell+\hlf}^2+
    \frac{1}{2\mu_0}\sum_\ell\Delta x B_{\perp,\ell+\hlf}^2+
    \sum_p\frac{w_pm_p}{2}v_{p,x}^2,
\end{equation}
and $\bar{v}_s\equiv\sqrt{\sum_{p}v_{s,x,p}^2}$ is the species's characteristic velocity. As observed in previous studies and due to the stringent tolerances imposed on the HO Picard convergence criterion in this test case, continuity and canonical momentum are conserved to numerical precision, while energy is conserved to within the HOLO tolerance $\left({\rm tol}_{\HO\LO} = 10^{-8}\right)$. Although linear momentum is not conserved exactly, the associated error is controlled through physics-based substep constraints, as discussed in~\S\ref{sec:HO_discrete}. \par

\paragraph{\textbf{Solver Performance}}Table~\ref{tab:EW_HOLO} presents a comparison of the average number of HOLO iterations per timestep for the electron Weibel instability, evaluated across various timestep sizes and LO systems. 

\begin{table}[h!]
    \centering
    \caption{Electromagnetic electron Weibel instability: Average HOLO iterations per timestep for the electron Weibel instability, comparing three LO systems with $N_{\ppc} = 3000$.}
    \def\arraystretch{1.4}
    \begin{tabular}{|c||c|c|c|c|}
         \hline
         $\Delta t\omega_{p\e}$&10&50&100&200  \\
         \hhline{|=||=|=|=|=|}
         4MP& 24.58 & 132.4 & 197.9 & 198.7\\
         \hline
         5MP& 23.36 & 123.4 & 133.1 & 158.2\\
         \hline
         7MP& 23.26 & 123.6 & 135.2 & 166.0\\
         \hline
    \end{tabular}
    \label{tab:EW_HOLO}
\end{table}

As with previous test cases, for small timestep sizes ($\Delta t\omega_{\p\e} \leq 10$), the performance of the different LO models is nearly identical. However, for larger $\Delta t$, the 5-moment system requires up to approximately 30\% fewer HOLO iterations than the 4-moment system. The number of HOLO iterations increases substantially as the timestep grows, primarily due to the finite number of particles and the stringent solver tolerances, as further illustrated below.\par
Figure~\ref{fig:EW_Error} shows the convergence behavior of the HOLO residual ($\|\mathbf{r}\|$, defined in Eqn.~\eqref{eq:HOLO_res}) as a function of HOLO iteration count at a particular representative timestep, varying $N_{ppc}$, $\Delta t$ and the LO system. Note that Table~\ref{tab:EW_HOLO} and Fig.~\ref{fig:EW_Error} use different $N_{\ppc}$ and Table~\ref{tab:EW_HOLO} shows time averaged data while Fig.~\ref{fig:EW_Error} only shows results at a single timestep, so the data are not directly comparable. 

\begin{figure}[h!]
    \centering
    \includegraphics[width=450pt]{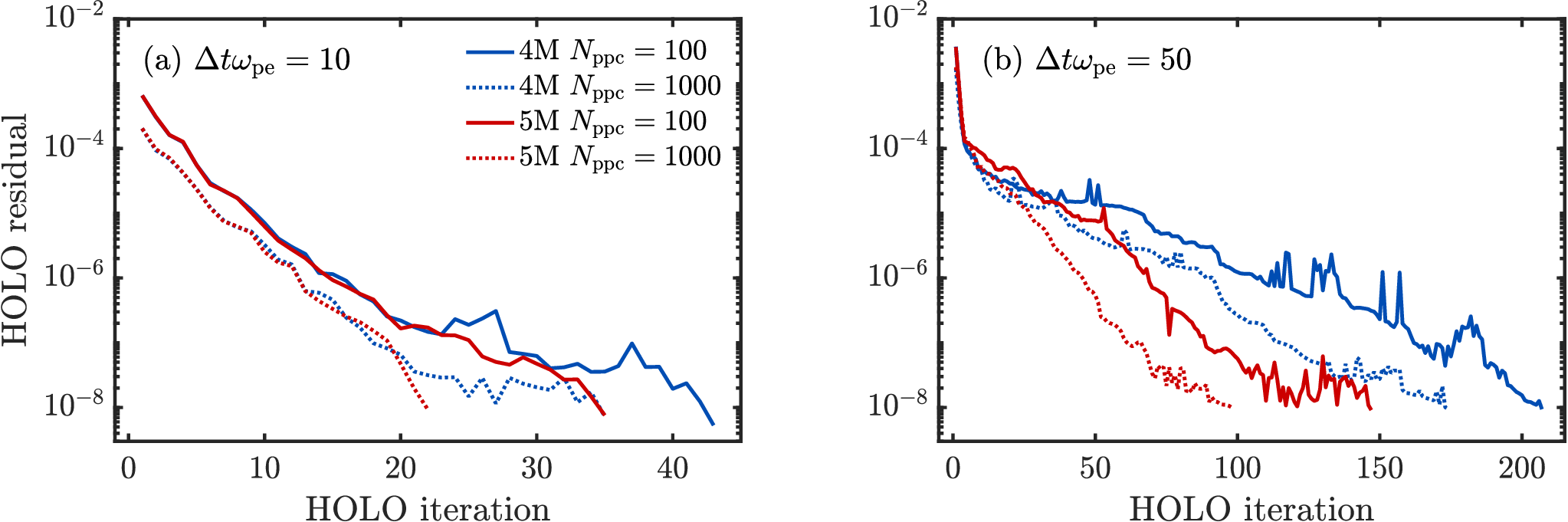}
    \caption{Electromagnetic electron Weibel instability: HOLO residual $\|\mathbf{r}\|$ (as defined in Algorithm~\ref{alg:HOLO}) versus HOLO iteration number at a particular representative timestep. Shown are the results using the primitive 4-moment (4M) and 5-moment (5M) systems, using $N_{\ppc} = 100$ and $1000$, for (a) $\Delta t\omega_{\p\e} = 10$ and (b) $\Delta t\omega_{\p\e} = 50$.}
    \label{fig:EW_Error}
\end{figure}

Results are presented for the primitive 4- and 5-moment LO systems, with $N_{\ppc} \in \{100, 1000\}$, and timestep sizes $\Delta t\omega_{\p\e} \in \{10, 50\}$. For $\Delta t\omega_{\p\e} = 10$, a clear distinction is observed between the two particle counts. The case with $N_{\ppc} = 1000$ starts from a lower initial residual and converges at a similar rate, requiring fewer HOLO iterations. This illustrates the role of particle noise in influencing solver convergence. Additionally, while the 4- and 5-moment LO systems initially behave similarly, the 4-moment system’s residual levels off and converges more slowly, whereas the 5-moment system continues to reduce the residual more rapidly.

These results suggest that if a looser HOLO tolerance (corresponding to a looser energy conservation tolerance) were permissible, the differences in iteration count between the LO systems would shrink. Conversely, for tight tolerances, the 5-moment system shows clear advantages.\par

At $\Delta t\omega_{\p\e} = 50$, the behavior changes. All test cases initially exhibit rapid convergence for the first $\sim 10$ iterations, reaching $\|\mathbf{r}\| \approx 10^{-4}$. Beyond that, each model enters an approximately exponential decay phase, with decay rates dependent on the LO system. For both $N_{\ppc} = 100$ and $1000$, the 5-moment system converges more quickly than the 4-moment system, requiring fewer HOLO iterations. While the convergence rate is relatively insensitive to particle count, larger $N_{\ppc}$ results in smaller residual fluctuations, as seen in Fig.~\ref{fig:EW_Error}(b).\par

\begin{table}[h!]
    \centering
    \caption{Electromagnetic electron Weibel instability: Solver statistics for the electron Weibel instability with the 5-moment primitive LO system with $N_{\ppc}=3000$.}
    \def\arraystretch{1.4}
    \begin{tabular}{|c||c|c|c|c|}
    \hline
     $\Delta t\omega_{\p\e}$&10&50&100&200  \\
     \hhline{|=||=|=|=|=|}
     $\left(\frac{\textrm{Picard itns.}}{\textrm{Substep}}\right)_{\rm avg}$& 3.5&3.7&4.1&4.6\\
     \hline
     $\left(\frac{\textrm{Particle substeps}}{\textrm{HOLO itn.}}\right)_{\rm avg}$& 1.3&3.2&5.6&10.8\\
     \hline
     $\left(\frac{\textrm{LO itns.}}{\textrm{HOLO itn.}}\right)_{\rm avg}$& 5.2&9.6&13.4&12.6\\
     \hline
     $\left(\frac{\textrm{HOLO itns.}}{\textrm{Timestep}}\right)_{\rm avg}$& 23.36&123.4&133.1&158.2\\
     \hline
     Runtime& 1 (16276 s)&1.609&1.062&0.715\\
     \hline
\end{tabular}\label{tab:EW_Solver}
\end{table}

\paragraph{\textbf{Solver Statistics}}Solver statistics for the 5-moment primitive LO model are shown in Table~\ref{tab:EW_Solver}. Despite the use of a conservative LO preconditioner with a primitive residual, the number of LO iterations increases slowly (sub-linearly) across a wide range of timestep sizes.

\begin{figure}[h!]
    \centering
    \includegraphics[width = \linewidth]{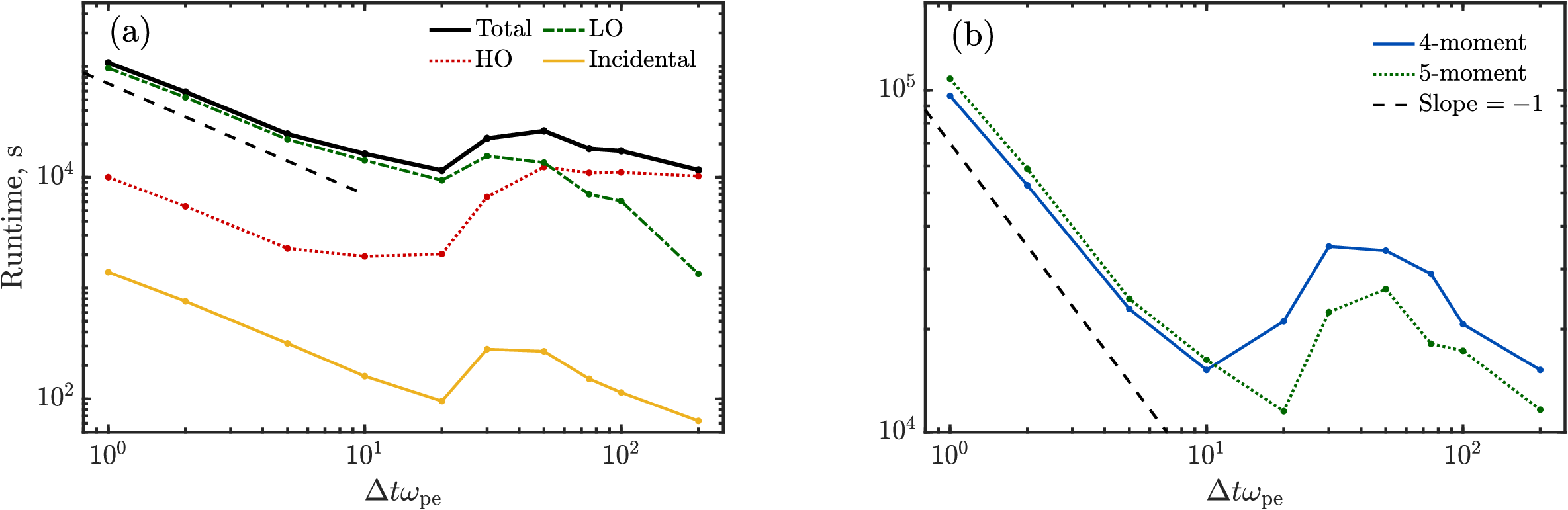}
    \caption{Electromagnetic electron Weibel instability with $N_{\ppc}=3000$: (a) Subroutine-level runtime breakdown for the 5-moment LO system. (b) Comparison of total runtime between 4- and 5-moment LO systems as a function of timestep size.}
    \label{fig:EWRun}
\end{figure}

Similarly, as shown in Table~\ref{tab:EW_HOLO}, for large timestep sizes ($\Delta t\omega_{\p\e} \gtrsim 100$), the 5-moment system requires significantly fewer HOLO iterations than the 4-moment system, leading to a marked reduction in overall runtime. Figure~\ref{fig:EWRun}(a) provides a subroutine-level runtime breakdown for the 5-moment system, while Fig.~\ref{fig:EWRun}(b) compares the total runtime between the 4- and 5-moment LO systems. At small timestep sizes ($\Delta t\omega_{\p\e} < 10$), runtime decreases nearly linearly with increasing $\Delta t$ for both LO models. This is expected since the dynamical timescale ($\omega_\mathrm{I}^{-1}\omega_{\p\e} \simeq 250$) allows convergence in just a few iterations, with the linear speedup arising from taking fewer timesteps. Figure~\ref{fig:EWRun}(a) shows that the LO system dominates the runtime cost for small timestep sizes, $\Delta t\omega_{\p\e}\lesssim10$, and   $N_{\ppc=3000}$. At small timestep sizes, particles only require 1--2 substeps, and converge quickly since the field does not change much over the particles' short path. Meanwhile, the LO system must still capture the exponential growth of the magnetic field, which can be costly. Our LO preconditioner neglects the magnetic couplings between the fluid quantities, possibly contributing to slow LO convergence for electromagnetic problems. Table~\ref{tab:EW_Solver} shows that for $\Delta t\omega_{\p\e} =10$, there is an average of $5.2$ LO iterations per HOLO iteration, which is greater than all of the Landau damping test cases (cf. Table~\ref{tab:LD_Solver}).\par
For intermediate timesteps ($10 < \Delta t\omega_{\p\e} < 50$), the total runtime increases, mainly due to growth in the runtime of the HO system, as the fastest particles begin requiring frequent substepping. This regime also sees an increase in the HOLO iteration count, reflecting more stringent coupling demands between HO and LO systems.  From Figure~\ref{fig:EWRun}(a), we can see that at this point the computational cost of the HO system increases with $\Delta t$, due to the requirement of increased substepping. At  larger timestep sizes ($\Delta t\omega_{\p\e} > 50$), the total runtime stabilizes and even decreases. This is because particle Picard convergence remains rapid, and particles are already being pushed across entire cell widths, leading to subcycling saturation. Meanwhile, the number of LO iterations per HOLO iteration only increases modestly (from $9.6$ to $12.6$). Hence, the runtime becomes primarily limited by HOLO iterations, which do not scale linearly with timestep, offering further efficiency gains at large $\Delta t$.\par

As Fig.~\ref{fig:EWRun}(b) shows, at small timesteps ($\Delta t\omega_{\p\e} \lesssim 10$), the 5-moment system incurs slightly higher computational cost due to the increased complexity of the LO solve. However, for larger timestep sizes ($\Delta t\omega_{\p\e} \gtrsim 10$), the more physically accurate linear dispersion of the 5-moment system allows it to converge in fewer HOLO iterations, offsetting the additional cost of its LO solver. Furthermore, as timestep size increases, the proportion of runtime spent on the LO system diminishes, with the HO solve dominating the total cost. Thus, the reduction in HOLO iterations offered by the 5-moment system offsets and outweighs the increased complexity of the LO solver.

\subsection{Electromagnetic ion Weibel instability}

\paragraph{\textbf{Verification}}The ion Weibel instability is selected as a complementary test case to the electron Weibel problem, with dynamics now occurring on the ion timescale. Since we still simulate electrons, explicit schemes would still require resolution of electron-scale dynamics for stability; however, we demonstrate that the electromagnetic implicit HOLO algorithm can stably advance the solution over timescales on the order of $\omega_{\p\ii}^{-1}$.\par

The dispersion relation for the ion Weibel instability is identical to that given in Eqn.~\eqref{eq:WeibelDisp}, but with the temperature anisotropy present in the ion VDF. For this test, in comparison with Ref.~\cite{chen2014energy}, we use a periodic domain of length $L_x = 2\pi/\left(3\sqrt{1836}\right)$, with $N_x = 32$ and $N_{\ppc} = 3000$.\par

Electrons are initialized as a hot, uniform, isotropic Maxwellian plasma with $v_{\e,\textrm{th},x} = v_{\e,\textrm{th},\perp} = 0.025$, $n_\e(x;t=0) = 1$, and $u_{\e,i}(x;t=0) = 0$. The ions are initialized as a warm, anisotropic Gaussian distribution with $v_{\ii,\textrm{th},x} = 0.001$ and $v_{\ii,\textrm{th},\perp} = 0.2$, $n_\ii(x;t=0) = 1$, and $u_{\ii,i}(x;t=0) = 0$. This strong ion temperature anisotropy, $T_\perp / T_x = 4 \times 10^4$, is chosen to yield a large ion Weibel growth rate of $\omega_\mathrm{I} = 2.4 \times 10^{-3}\ \omega_{\p\e}^{-1} = 0.102\ \omega_{\p\ii}^{-1}$. The simulation is run until $t\omega_{\p\ii} = 100$.\par

\begin{figure}[h!]
    \centering
    \includegraphics[width=300pt]{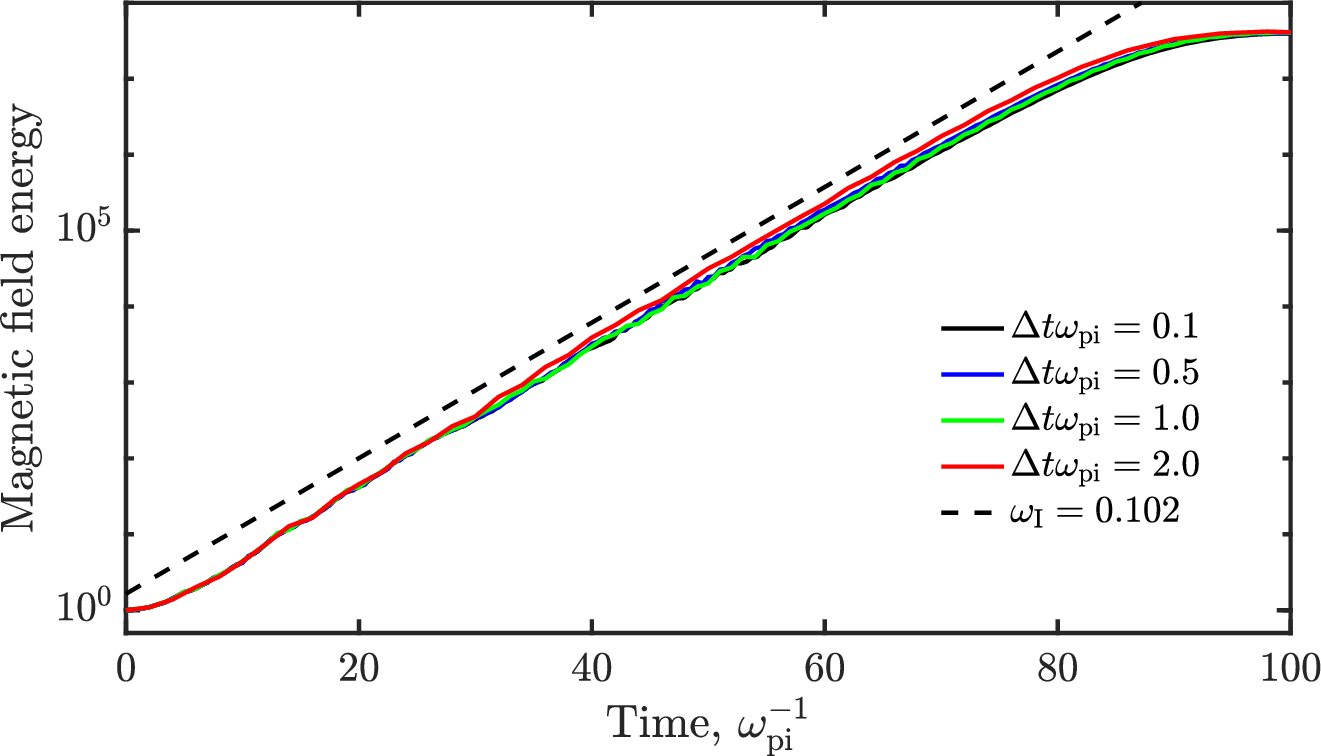}
    \caption{Electromagnetic ion Weibel instability: Plot of the magnetic field energy as a function of time, comparing solutions obtained with various timestep sizes to the analytical growth rate. These results use the 4-moment primitive formulation of the LO system with $N_{\ppc} = 3000$.}
    \label{fig:IW_NRG}
\end{figure}

Figure~\ref{fig:IW_NRG} presents the results of the ion Weibel instability test case, showing excellent agreement with the theoretical linear growth rate across all chosen values of $\Delta t$. As with previous cases, we demonstrate that the HOLO algorithm can accurately resolve linear growth dynamics for timesteps up to $\Delta t = 2 \omega^{-1}_{\p\ii} = 0.2 \omega^{-1}_\mathrm{I}$, substantially larger than the explicit stability limit and approaching that of the dynamical scale (i.e., the growth-time scale). \par 

\begin{table}[h!]
    \centering
    \caption{Electromagnetic ion Weibel instability: Average HOLO iterations per timestep, comparing three LO systems with $N_{\ppc} = 3000$.}
    \def\arraystretch{1.4}
    \begin{tabular}{|c||c|c|c|c|}
        \hline
         $\Delta t\omega_{\p\ii}\textrm{ }(\Delta t\omega_{\p\e})$&0.1\textrm{ }(4.3)&0.5\textrm{ }(21.4)&1.0\textrm{ }(42.8)&2.0\textrm{ }(85.7)  \\
         \hhline{|=||=|=|=|=|}
         4MP& 6.59 & 8.01 & 13.31 & 50.68\\
         \hline
         5MP& 5.31 & 7.46 & 11.42 & 45.94\\
         \hline
         7MP& 5.29 & 7.36 & 11.99 & 46.02\\
         \hline
    \end{tabular}
    \label{tab:IW_HOLO}
\end{table}

\paragraph{\textbf{Solver Performance and Statistics}} Table~\ref{tab:IW_HOLO} summarizes the average number of HOLO iterations per timestep for different values of $\Delta t$ and LO systems. As seen in previous studies, the 5- and 7-moment systems consistently require fewer HOLO iterations than the 4-moment system, showing improvements of approximately 10--25\% across all timestep sizes. However, unlike earlier cases, the number of HOLO iterations does not decrease monotonically with increasing $\Delta t$; in fact, it grows super-linearly. This behavior suggests a practical upper limit on timestep size for the chosen LO systems. While larger timesteps reduce the number of particle pushes and thus, in principle, reduce overall computational cost, the increasing difficulty of converging the coupled HOLO system at large $\Delta t$ diminishes these benefits.\par 

\begin{table}[h!]
    \centering
    \caption{Electromagnetic ion Weibel instability: Solver statistics with the 5-moment primitive LO system with $N_{\ppc} = 3000$}
    \def\arraystretch{1.4}
    \begin{tabular}{|c||c|c|c|c|}
    \hline
     $\Delta t\omega_{\p\ii}$&0.1&0.5&1.0&2.0  \\
     \hhline{|=||=|=|=|=|}
     $\left(\frac{\textrm{Picard itns.}}{\textrm{Substep}}\right)_{\rm avg}$& 4.0&6.0&7.9&8.5\\
     \hline
     $\left(\frac{\textrm{Particle substeps}}{\textrm{HOLO itn.}}\right)_{\rm avg}$& 3.9&16.1&31.0&59.6\\
     \hline
     $\left(\frac{\textrm{LO itns.}}{\textrm{HOLO itn.}}\right)_{\rm avg}$& 2.2&2.2&2.2&2.3\\
     \hline
     $\left(\frac{\textrm{HOLO itns.}}{\textrm{Timestep}}\right)_{\rm avg}$& 5.3&7.5&11.4&45.9\\
     \hline
     Runtime& 1 (13569 s)&0.743&1.417&4.365\\
     \hline
\end{tabular}\label{tab:IW_Solver}
\end{table}

These trends are further supported by the solver statistics in Table~\ref{tab:IW_Solver}. For $\Delta t\omega_{\p\ii} = 2$, there is a notable increase in both the number of substeps per particle (nearly doubled) and the number of HOLO iterations (nearly quadrupled), resulting in an almost threefold increase in total runtime. This indicates that for larger timesteps, the LO solver, which under-represents ion dynamics, struggles to provide accurate field estimates, thereby degrading HOLO convergence efficiency. \par

\begin{figure}[h!]
    \centering
    \includegraphics[width=450pt]{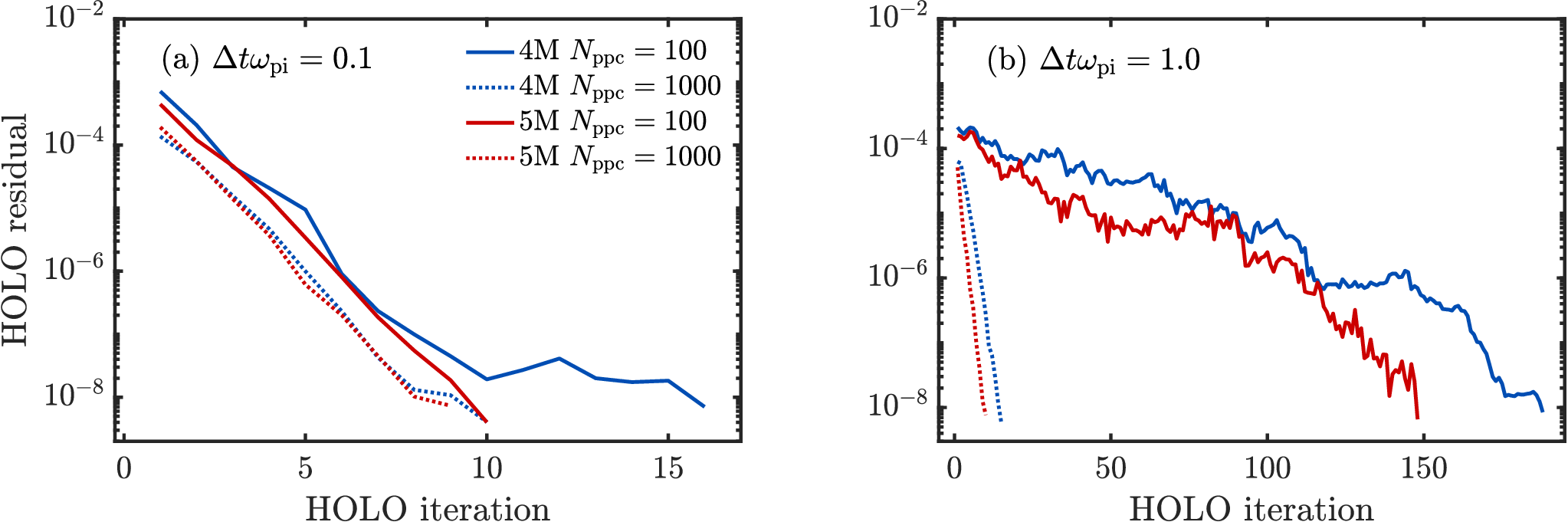}
    \caption{Electromagnetic ion Weibel instability: HOLO residual $\|\mathbf{r}\|$ (as defined in Algorithm~\ref{alg:HOLO}) versus HOLO iteration number at a particular representative timestep. Shown are the results using the primitive 4-moment (4M) and 5-moment (5M) systems, using $N_{\ppc} = 100$ and $1000$, for (a) $\Delta t\omega_\mathrm{pi} = 0.1$ and (b) $\Delta t\omega_\mathrm{pi} = 1.0$.}
    \label{fig:IW_Error} 
\end{figure}

Figure~\ref{fig:IW_Error} shows the convergence behavior of the HOLO residual as a function of HOLO iteration count at a particular representative timestep, varying $N_{\ppc}$, $\Delta t$ and the LO system. As before, note that Table~\ref{tab:IW_HOLO} and Fig.~\ref{fig:IW_Error} use different $N_{\ppc}$ and Table~\ref{tab:IW_HOLO} shows time-averaged data while Fig.~\ref{fig:IW_Error} only shows results at a single timestep, so the data are not directly comparable. Results are presented for the primitive 4- and 5-moment LO systems, with $N_{\ppc} \in [100, 1000]$ and $\Delta t\omega_{\p\ii} \in [0.1, 1.0]$. In Fig.~\ref{fig:IW_Error}(a), corresponding to $\Delta t\omega_{\p\ii} = 0.1$, the timestep is smaller than the dynamical scale ($\Delta t\omega_\mathrm{I} \simeq 0.01$), and the residual generally decreases exponentially. However, for the 4-moment case with $N_{\ppc} = 100$, convergence stalls slightly around $10^{-8}$, requiring approximately 50\% more HOLO iterations than the 5-moment model. Increasing $N_{\ppc}$ improves residual behavior marginally by mitigating particle noise. \par

The impact of particle statistics becomes much more significant at $\Delta t\omega_{\p\ii} = 1.0$, as shown in Fig.~\ref{fig:IW_Error}(b). The case with $N_{\ppc} = 100$ requires over 10 times more HOLO iterations than $N_{\ppc} = 1000$, highlighting how particle noise, poorly represented in moment-based models, can severely impair convergence. This demonstrates that increasing the number of particles can, in some cases, actually reduce total runtime. Additionally, the 5-moment model consistently maintains a lower residual and faster convergence rate compared to the 4-moment system, further reducing the number of required HOLO iterations.\par

\begin{figure}[h!]
    \centering
    \includegraphics[width=0.8\linewidth]{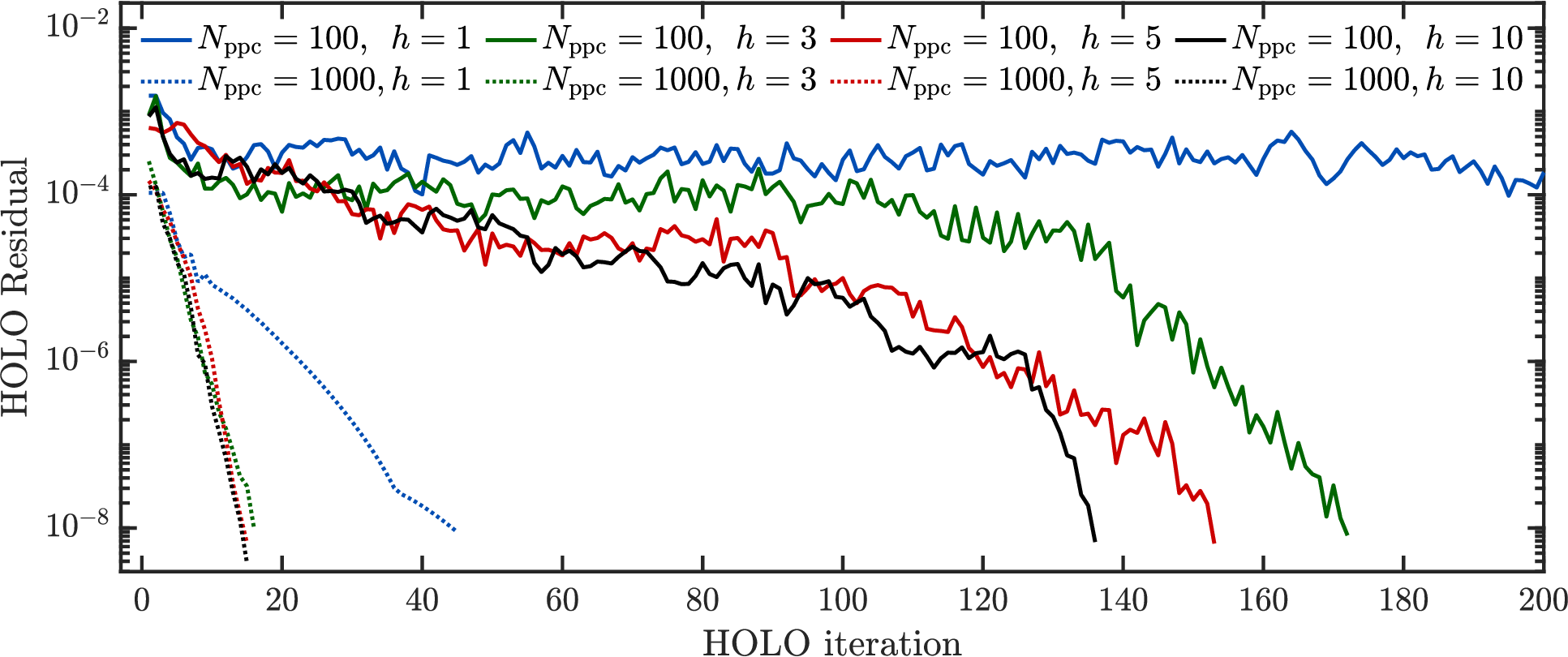}
    \caption{Electromagnetic ion Weibel instability: HOLO residual convergence as a function of iteration count, for varying particle numbers $N_{\ppc} = \left\{100, 1000 \right\}$ and Anderson mixing history lengths $h = \left\{1, 3, 5, 10 \right\}$.}
    \label{fig:aa_impact_on_convergence}
\end{figure}

Finally, we report that the nonlinear convergence of the HOLO solver can be significantly accelerated by employing Anderson mixing. In Figure~\ref{fig:aa_impact_on_convergence}, we illustrate the sensitivity of HOLO residual convergence to variations in particle number and Anderson mixing history length for a timestep size of $\Delta t \omega_{\p\ii} = 1.0$.\par

As shown in Fig.~\ref{fig:aa_impact_on_convergence}, for a low number of particles per cell ($N_{\ppc} = 100$), particle noise can significantly degrade and even stall the convergence of the HOLO solver in the absence of Anderson mixing (i.e., $h = 1$). Consistent with the observations in Fig.~\ref{fig:IW_Error}, increasing the number of particles, i.e, reducing the particle noise, yields the greatest reduction in the number of iterations required. Even in the high-particle limit, we found that even a small amount of Anderson mixing ($h=3$) \textit{halved} the number of HOLO iterations required. Importantly, this improvement is achieved at relatively low cost, as Anderson mixing requires only the storage of a few moments and field quantities, so the addition of Anderson mixing effectively halved the simulation runtime. However, we observed that the marginal benefit of Anderson mixing diminished beyond $h = 3$ and there was an insignificant speedup associated with employing $h>5$.
\par 
The ability of Anderson mixing to enhance solver robustness without increasing particle count is especially valuable in extreme-scale simulations, where computational resources are typically utilized to their limits. Finally, we note that convergence behavior can vary significantly from timestep to timestep, particularly in low-particle cases. All simulations shown were performed using the same initial random seed and compared at a common time, later in the instability growth phase, $t = 50\omega_{\p\ii}^{-1}$. The motivation for choosing this point in time is that as the instability grows and excites stronger fields, the number of HOLO iterations tends to grow. This time roughly yielded the number of HOLO iterations that corresponded to the average value over the entire simulation for the setup shown in Table \ref{tab:IW_HOLO} using the 5-moment solver (roughly 11.42) and therefore used as the reference for comparison. 

\section{Conclusions}\label{sec:Conclusions}

This study has extended previous work on improved HOLO moment-based accelerators by incorporating them into a charge- and energy-conserving electromagnetic framework. We compared the performance of different moment systems and found that the HOLO approach achieved robust convergence for timestep sizes approaching the dynamical timescale of the system. Moreover, we observed that the structure of the moment system, and the physical waves that are permitted in its dispersion relation, can significantly influence the convergence behavior of the HOLO scheme. These findings pave the way for a more detailed investigation into how the choice of moment system can enhance the efficiency of implicit moment methods. In particular, moment systems that encapsulate a richer set of system dynamics may yield improved HOLO convergence performance. \par

This study focused on applying the model to physical setups with linear instability or decay rates. Future work may study the efficacy of the model in the nonlinear regime where the low-order fluid moment system may struggle to converge to the true kinetic solution. We also found that statistical noise from a finite number of particles can greatly deteriorate the convergence properties of the model, even with low-discrepancy sampling for the initial condition and binomially smoothed moment calculations. Subsequent investigations may study techniques to mitigate numerical noise in moment calculations such as variance reduction techniques to improve the performance of the HOLO solver. Future work may also involve extending this method to multiple spatial dimensions by generalizing the LO system to include transport terms in additional directions. Given that magnetic fields can induce anisotropic pressures and are generally not aligned with coordinate axes, it is hypothesized that moment systems capable of capturing a full pressure tensor (e.g., the 10-moment system in three dimensions) may outperform the traditional 5-moment (isotropic pressure) system in multidimensional settings.

\section*{Acknowledgements}
The authors thank Luis Chacon and Guangye Chen for many helpful discussions. D.A.K. was supported by a NASA Space Technology Graduate Research Opportunity, Grant No. 80NSSC21K1302, and the U.S. Department of Energy National Nuclear Security Administration Stewardship Science Graduate Fellowship under cooperative agreement DE-NA0003960. W.T.T was supported by Triad National Security, LLC under contract 89233218CNA000001 and the DOE Office of Applied Scientific Computing Research (ASCR) through the Mathematical Multifaceted Integrated Capability Centers program. The authors would like to thank Stanford University and the Stanford Research Computing Center for providing computational resources and support that contributed to these research results. 

\appendix
\vspace{10pt}

\section{Dispersion relations of Low-Order systems with High-Order closure}\label{sec:AppDisperion}
In this section, we present the derivation of the dispersion relations presented in \S\ref{sec:HOLO_LO}.
\subsection{4-moment system}
Consider Eqn.~\ref{eq:4MCP}; let us assume that  $n$ and $\Gamma_i$ take the form $Q = Q_0+Q'\exp\left[-i(\omega t-kx)\right]$ with $|Q_0|\ll \left|Q'\right|$. Doing so, we obtain the following coupled equations:
\begin{equation}
\begin{split}
    \textrm{Conservative: } &
    \begin{cases}
        \mathlarger{-i\omega n'+ik\Gamma_x'=0}\\
        \\
        \mathlarger{-i\omega \Gamma_i' +ikn'\tilde{S}_{xi}^\HO=0}\\
    \end{cases}
    \\
    \textrm{Primitive: }&
    \begin{cases}
        \mathlarger{-i\omega n'+ik\Gamma_x'=0}\\
        \\
        \mathlarger{-i\omega \Gamma_i' +ik\left(n'T_{xi}^\HO+\frac{\Gamma_{0,i}}{n_0}\Gamma_x'+\frac{\Gamma_{0,i}}{n_0}\Gamma_i'-\frac{\Gamma_{0,i}\Gamma_{0,x}}{n_0^2}n'\right)=0}\\
    \end{cases}
    \end{split}
\end{equation}
To solve these dispersion relations, we set up the system of linear equations as a matrix equation: $A\vec{\mathcal{M}}=0$, where $A$ is a coefficient matrix and $\vec{\mathcal{M}} = \left[n',\Gamma_x',\Gamma_y',\Gamma_z'\right]^T$ is the vector of moment quantities. Substituting $u_{0,i} = \Gamma_{0,i}/n_0$ and removing the $\HO$ superscript:
\begin{multline}
    \textrm{Conservative: }
    A_C=\begin{bmatrix}
        -i\omega & ik & 0 & 0\\
        ik\tilde{S}_{xx} & -i\omega & 0 & 0\\
        ik\tilde{S}_{xy} & 0 & -i\omega & 0\\
        ik\tilde{S}_{xz} & 0 & 0& -i\omega\\
    \end{bmatrix}
    \\
    \textrm{Primitive: }
    A_P=\begin{bmatrix}
        -i\omega & ik & 0 & 0\\
        ik\left(T_{xx}-u_{0,x}^2\right) & -i\omega+2iku_{0,x} & 0 & 0\\
        ik\left(T_{xy}-u_{0,x}u_{0,y}\right) & iku_{0,y} & -i\omega+iku_{0,x} & 0\\
        ik\left(T_{xz}-u_{0,x}u_{0,z}\right) & iku_{0,z} & 0& -i\omega+iku_{0,x}\\
    \end{bmatrix}
\end{multline}
Setting the determinant of the coefficient matrices to zero, the above equations have the following characteristic polynomials:
\begin{equation}
    \textrm{Conservative: }
    \omega^2\left(\omega^2-k^2\tilde{S}_{xx}\right)=0,\hspace{20pt}
    \textrm{Primitive: }
    \left(\omega-ku_x\right)^2\left[\left(\omega-ku_x\right)^2-k^2T_{xx}\right]=0.
\end{equation}
Solving these, we find the eigenvalues
\begin{equation}
    \textrm{Conservative: }
    \frac{\omega}{k} = \left\{0,0,\pm\tilde{S}_{xx}\right\},\hspace{20pt}
    \textrm{Primitive: }
    \frac{\omega}{k}-u_x = \left\{0,0,\pm\sqrt{T_{xx}}\right\},
\end{equation}
as presented in Eqn.~\eqref{eq:4M_Disp}.
\subsection{5-moment system}
Linearizing the equations in Eqn.~\ref{eq:5MCP} as in the previous section, and neglecting the electromagnetic source terms to focus on the dispersion relation of the fluid transport equations, we find:
\begin{equation}
\begin{split}
    \textrm{Conservative: }&
    \begin{cases}
        \mathlarger{-i\omega n'+ik\Gamma_{x}' = 0,}\\
        \\
        \mathlarger{-i\omega\Gamma_{x}+ikS_{xx}' = 0,}\\
        \\
        \mathlarger{-i\omega\Gamma_{\perp}+ikn'\tilde{S}^\HO_{x\perp} = 0,}\\
        \\
        \mathlarger{-i\omega S_{xx}'+ikn' \tilde{Q}^\HO_{s,xxx} = 0,}
    \end{cases}
\\
\textrm{Primitive: }&
    \begin{cases}
    \mathlarger{-i\omega n'+ik\Gamma_{x}' = 0,}\\
    \\
    \mathlarger{-i\omega\Gamma_{x}+ikS_{xx}' = 0,}\\
    \\
    \mathlarger{-i\omega \Gamma_\perp' +ik\left(n'T_{x\perp}^\HO+\frac{\Gamma_{0,\perp}}{n_0}\Gamma_x'+\frac{\Gamma_{0,\perp}}{n_0}\Gamma_\perp'-\frac{\Gamma_{0,i}\Gamma_{0,x}}{n_0^2}n'\right)=0,}\\
    \\
    \mathlarger{-i\omega S_{xx}'+ik\left[n'\tilde{q}^\HO_{s,xxx}
    +\frac{3S_{0,xx}}{n_0}\Gamma_x' 
    +\frac{3\Gamma_{0,x}}{n_0}S_{xx}'
    -\frac{3\Gamma_{0,x}S_{0,xx}}{n_0^2}n'
    -\frac{6\Gamma_{0,x}^2}{n_0^2}\Gamma_x'
    +\frac{4\Gamma_{0,x}^3}{n_0^3}n'\right] = 0,}
\end{cases}
\end{split}
\end{equation}
To solve these dispersion relations, we set up the system of linear equations as a matrix equation: $A\vec{\mathcal{M}}'=0$, where $A$ is a coefficient matrix and $\vec{\mathcal{M}}' = \left[n',\Gamma_x',\Gamma_y',\Gamma_z',S_{xx}'\right]^T$ is the vector of moment quantities. Substituting $u_{0,i} = \Gamma_{0,i}/n_0$ and $T_{0,xx} = S_{0,xx}/n_0-\Gamma_{0,x}^2/n_0^2$ and removing the $\HO$ superscript:
\begin{multline}
    \textrm{Conservative: }
    A_C=\begin{bmatrix}
        -i\omega & ik & 0 & 0 & 0\\
        0 & -i\omega & 0 & 0 & ik\\
        ik\tilde{S}_{xy} & 0 & -i\omega & 0 & 0\\
        ik\tilde{S}_{xz} & 0 & 0& -i\omega & 0\\
        ik\tilde{Q}_{xxx} & 0 & 0& 0 & -i\omega\\
    \end{bmatrix}
    \\
    \textrm{Primitive: }
    A_P=\begin{bmatrix}
        -i\omega & ik & 0 & 0 & 0\\
        0 & -i\omega+2iku_{0,x} & 0 & 0 & ik\\
        ik\left(T_{xy}-u_{0,x}u_{0,y}\right) & iku_{0,y} & -i\omega-iku_{0,x} & 0 & 0\\
        ik\left(T_{xz}-u_{0,x}u_{0,z}\right) & iku_{0,z} & 0& -i\omega+iku_{0,x} & 0\\
        ik\left(\tilde{q}_{xxx}+u_{0,x}^3-3u_{0,x}T_{0,xx}\right) & 3ik\left(T_{0,xx}-u_{0,x}^2\right) & 0& 0 & -i\omega+3iku_{0,x}\\
    \end{bmatrix}.
\end{multline}
Setting the determinant of the coefficient matrices to zero, the above equations have the following characteristic polynomials:
\begin{multline}
    \textrm{Conservative: }
    \omega^2\left(\omega^3-k^3\tilde{Q}_{xxx}\right)=0,\\
    \textrm{Primitive: }
    \left(\omega-ku_{0,x}\right)^2\left\{\left(\omega-ku_{0,x}\right)\left[\left(\omega-ku_{0,x}\right)^2-3k^2T_{xx}\right]-k^3\tilde{q}_{xxx}\right\}=0,
\end{multline}
Solving these in the limit of $\tilde{q}_{xxx}\rightarrow0$, we find the eigenvalues
\begin{equation}
    \textrm{Conservative: }
    \frac{\omega}{k} = \left\{0,0,\sqrt[3]{\tilde{Q}_{xxx}},e^{\pm2\pi i/3}\sqrt[3]{\tilde{Q}_{xxx}}\right\},\hspace{20pt}
    \textrm{Primitive: }
    \frac{\omega}{k} -u_x = \left\{0,0,0,\pm\sqrt{3T_{xx}}\right\}.
\end{equation}
To find how the heat flux affects the primitive solutions, let us define $c = \omega/k-u_x$, then the primitive characteristic equation can be written as
\begin{equation}\label{eq:B9}
    c\left(c^2-3T_{xx}^2\right)-\tilde{q}_{xxx}=f(c)-\tilde{q}_{xxx}=0,
\end{equation}
where we have eliminated the two roots at $c=0$. Although this cubic equation can be solved exactly, it is more illuminating to evaluate the wave speeds assuming that $\tilde{q}_{xxx}$ is small. $f(c)$ is approximately linear in the neighborhood of its roots so, assuming $\tilde{q}_{xxx}$ is small, Eqn.~\ref{eq:B9} has solutions
\begin{equation}
    c_1 = 0+\frac{\tilde{q}_{xxx}}{\left[df/dc\right]_{c=0}},\hspace{10pt}
    c_2 = \sqrt{3T_{xx}}+\frac{\tilde{q}_{xxx}}{\left[df/dc\right]_{c=\sqrt{3T_{xx}}}},\hspace{10pt}
    c_3 = -\sqrt{3T_{xx}}+\frac{\tilde{q}_{xxx}}{\left[df/dc\right]_{c=-\sqrt{3T_{xx}}}},
\end{equation}
and the full set of eigenvalues for the primitive equations is
\begin{equation}
    \textrm{Primitive: }
    \frac{\omega}{k} -u_x = \left\{0,0,-\frac{\tilde{q}_{xxx}}{3T_{xx}},\pm\sqrt{3T_{xx}}+\frac{\tilde{q}_{xxx}}{6T_{xx}}\right\},
\end{equation}
as presented in the main text.

\subsection{7-moment system}
Linearizing the equations in Eqn.~\ref{eq:7MC}
as in the previous section, and neglecting the electromagnetic source terms to focus on the dispersion relation of the fluid transport equations, we find:
\begin{equation}
\begin{split}
\textrm{Conservative: }&
\begin{cases}
    \mathlarger{-i\omega n'+ik\Gamma_{x}' = 0,}\\
    \\
    \mathlarger{-i\omega\Gamma_{i}+ikS_{ix}' = 0,}\\
    \\
    \mathlarger{-i\omega S_{ix}'+ikn' \tilde{Q}^\HO_{s,ixx} = 0,}
\end{cases}
\\
\textrm{Primitive: }&
    \begin{cases}
    \mathlarger{-i\omega n'+ik\Gamma_{x}' = 0,}\\
    \\
    \mathlarger{-i\omega\Gamma_{x}+ikS_{ix}' = 0,}\\
    \\
    \mathlarger{-i\omega S_{ix}'+ik\Bigg[n'\tilde{q}^\HO_{s,ixx}
    +\frac{2S_{0,ix}}{n_0}\Gamma_x' 
    +\frac{2\Gamma_{0,x}}{n_0}S_{ix}'
    -\frac{2\Gamma_{0,x}S_{0,ix}}{n_0^2}n'}\\
    \hspace{25pt}\mathlarger{+\frac{S_{0,xx}}{n_0}\Gamma_i' 
    +\frac{\Gamma_{0,i}}{n_0}S_{xx}'
    -\frac{\Gamma_{0,i}S_{0,xx}}{n_0^2}n'
    -\frac{4\Gamma_{0,i}\Gamma_{0,x}}{n_0^2}\Gamma_x'
    -\frac{2\Gamma_{0,x}^2}{n_0^2}\Gamma_i'
    +\frac{4\Gamma_{0,i}\Gamma_{0,x}^2}{n_0^3}n'\Bigg] = 0.}
\end{cases}
\end{split}
\end{equation}
To solve these dispersion relations, we set up the system of linear equations as a matrix equation: $A\vec{\mathcal{M}}=0$, where $A$ is a coefficient matrix and $\vec{\mathcal{M}} = \left[n',\Gamma_x',\Gamma_y',\Gamma_z'\right]$ is the vector of moment quantities. Substituting $u_{0,i} = \Gamma_{0,i}/n_0$ and $T_{0,xx} = S_{0,xx}/n_0-\Gamma_{0,x}^2/n_0^2$, defining $L_{ij} = T_{ij}-u_iu_j$ and removing the $\HO$ superscript:
\begin{multline}
    \textrm{Conservative: }
    A_C=\begin{bmatrix}
        -i\omega & ik & 0 & 0 & 0 & 0 & 0\\
        0 & -i\omega & 0 & 0 & ik & 0 & 0\\
        0 & 0 & -i\omega & 0 & 0 & ik & 0\\
        0 & 0 & 0& -i\omega & 0 & 0 & ik\\
        ik\tilde{Q}_{xxx} & 0 & 0& 0 & -i\omega & 0 & 0\\
        ik\tilde{Q}_{yxx} & 0 & 0& 0 & 0 & -i\omega & 0\\
        ik\tilde{Q}_{zxx} & 0 & 0& 0 & 0 & 0 & -i\omega\\
    \end{bmatrix}
    \\
    \textrm{Primitive: }\hspace{350pt}
    \\
    A_P=\begin{bmatrix}
        -i\omega & ik & 0 & 0 & 0 & 0 & 0\\
        0 & -i\omega & 0 & 0 & ik & 0 & 0\\
        0 & 0 & -i\omega & 0 & 0 & ik & 0\\
        0 & 0 & 0& -i\omega & 0 & 0 & ik\\
        ik\left(\tilde{q}_{xxx}-u_xL_{xx}-2u_{x}T_{xx}\right) & 3ikL_{xx} & 0& 0 & -i\omega+3iku_{x} & 0 & 0\\
        ik\left(\tilde{q}_{xxy}-u_yL_{xx}-2u_{x}T_{xy}\right) & 2ikL_{xy} & ikL_{xx}& 0 & 2iku_{x} & -i\omega+iku_{x} & 0\\
        ik\left(\tilde{q}_{xxz}-u_zL_{xx}-2u_{x}T_{xz}\right) & 2ikL_{xz} & 0& ikL_{xx} & 2iku_{x} & 0 & -i\omega+iku_{x}\\
    \end{bmatrix}.
\end{multline}
Setting the determinant of the coefficient matrices to zero, the above equations have the following characteristic polynomials:
\begin{multline}
    \textrm{Conservative: }
    \omega^4\left(\omega^3-k^3\tilde{Q}_{xxx}\right)=0,\\
    \textrm{Primitive: }
    \left[\left(\omega-ku_{0,x}\right)^2-k^2T_{xx}\right]^2\left\{\left(\omega-ku_{0,x}\right)\left[\left(\omega-ku_{0,x}\right)^2-3k^2T_{xx}\right]-k^3\tilde{q}_{xxx}\right\}=0.
\end{multline}
Solving these for small $\tilde{q}_{xxx}$ using the method in the previous section, we find the eigenvalues
\begin{multline}
    \textrm{Conservative: }
    \frac{\omega}{k} = \left\{0,0,0,0,\sqrt[3]{\tilde{Q}_{xxx}},e^{\pm2\pi i/3}\sqrt[3]{\tilde{Q}_{xxx}}\right\},\\
    \textrm{Primitive: }
    \frac{\omega}{k} -u_x = \left\{-\frac{\tilde{q}_{xxx}}{3T_{xx}},\pm\sqrt{T_{xx}},\pm\sqrt{T_{xx}},\pm\sqrt{3T_{xx}}+\frac{\tilde{q}_{xxx}}{6T_{xx}}\right\},
\end{multline}
which are the solutions presented in Eqn.~\eqref{eq:7M_Disp}.

\vspace{10pt}

\section{Shape functions and moment calculations} \label{sec:AppMoments}
\subsection{Shape Functions}
First-order shape functions $S_1$ are used to interpolate the electric field and gather the odd-$v_x$-order moments while second-order shape functions $S_2$ are used to gather even-$v_x$-order moments:
\begin{equation}\label{eq:S1}
    S_1(x,\Delta x) =
    \begin{cases}
        \mathlarger{1-\left|\frac{x}{\Delta x}\right|},& \textrm{for }\mathlarger{|x|\leq\Delta x}\\
        \\
        \mathlarger{0,} & \textrm{otherwise,}
    \end{cases}
\end{equation}
\begin{equation}\label{eq:S2}
    S_2(x,\Delta x) =
    \begin{cases}
        \mathlarger{\frac{3}{4}-\left|\frac{x}{\Delta x}\right|^2},& \textrm{for }\mathlarger{|x|\leq\frac{\Delta x}{2}}\\
        \\
        \mathlarger{\frac{1}{2}\left(\frac{3}{2}-\left|\frac{x}{\Delta x}\right|\right)^2},& \textrm{for }\mathlarger{\frac{\Delta x}{2}\leq|x|\leq\frac{3\Delta x}{2}}\\
        \\
        0, & \textrm{otherwise.}
    \end{cases}
\end{equation}
\subsection{Tracked moments}
The tracked moments (for which there is a moment equation) are gathered from particle positions, weights, and velocity, $\{x_p,w_p,v_p\}$ as: 
\begin{equation}
    n_\ell = \sum_p w_pS_2\left(x_p-x_\ell\right),
\end{equation}
\begin{equation}
    \Gamma_{x,\ell+\hlf} = \sum_p w_pv_{p,x}S_1\left(x_p-x_{\ell+\hlf}\right),\hspace{20pt}
    \Gamma_{\perp,\ell} = \sum_p w_pv_{p,\perp} S_2\left(x_p-x_{\ell}\right),
\end{equation}
\begin{equation}
    S_{xx,\ell} = \sum_p w_pv_{p,x}^2S_2\left(x_p-x_{\ell}\right),\hspace{20pt}
    S_{x\perp,\ell+\hlf} = \sum_p w_pv_{p,x}v_{p,\perp} S_1\left(x_p-x_{\ell+\hlf}\right).
\end{equation}
\subsection{Closure}
In this section, we present the exact form of how the High-Order closure quantities are calculated.
\subsubsection{4-moment conservative}
\begin{equation}
    \tilde{S}^{\eta+\hlf}_{xx,\ell} = \left(S_{xx,\ell}^{\eta+1}+S_{xx,\ell}^\eta\right)/\left(n_{\ell}^{\eta+1}+n_{\ell}^\eta\right)
\end{equation}
\begin{equation}
    \tilde{S}^{\eta+\hlf}_{x\perp,\ell+\hlf} = \left(S_{x\perp,\ell+\hlf}^{\eta+1}+S_{x\perp,\ell+\hlf}^\eta\right)/\left(n_{\ell+\hlf}^{\eta+1}+n_{\ell+\hlf}^\eta\right),
\end{equation}
recalling that $n_{\ell+\hlf} = \left(n_\ell+n_{\ell+1}\right)/2$. 
Calculated this way so that when multiplied by $n^{\eta+\hlf}$, we recover $S^{\eta+\hlf}$ exactly.

\subsubsection{4-moment primitive}
\begin{equation}
    T^{\eta+\hlf}_{xx,\ell} = \left[S_{xx,\ell}^{\eta+\hlf}-\left(\Gamma_{x,\ell}^{\eta+\hlf}\right)^2/n_{\ell}^{\eta+\hlf}\right]/n_{\ell}^{\eta+\hlf}
\end{equation}
and likewise for $T_{x\perp}$. Again, this is chosen to be calculated this way for consistency with the moment equations.
\subsubsection{5- and 7-moment conservative}

\begin{equation}
    Q_{xxx,\ell+\hlf} = \sum_p w_pv_{p,x}^3S_1\left(x_p-x_{\ell+\hlf}\right)
    \hspace{20pt}
    Q_{xx\perp,\ell} = \sum_p w_pv_{p,x}^2v_{p,\perp} S_2\left(x_p-x_{\ell}\right)
\end{equation}
\begin{equation}
    \tilde{Q} = Q/n
\end{equation}
\begin{equation}
    q_{xxi} = (Q_{xxi}-2\Gamma_xS_{xi}+\Gamma_iS_{xx}/n+2\Gamma_x^2\Gamma_i/n^2)/n
\end{equation}
\vspace{10pt}

\section{Discretized Low-Order equations} \label{sec:AppLO}

In this section we present the full forms of the discretized Low-Order equations.

\subsection{4-moment system}

This system is presented briefly in \S\ref{sec:Discretization} but reproduced in full here

\begin{equation}
    \frac{n^{\eta+1}_\ell-n^{\eta}_\ell}{\Delta t}+\frac{\Gamma_{x,\ell+\hlf}^{\eta+\hlf}-\Gamma_{x,\ell-\hlf}^{\eta+\hlf}}{\Delta x}=0,
\end{equation}
\begin{multline}
    \frac{\Gamma_{x,\ell+\hlf}^{\eta+\hlf}-\Gamma_{x,\ell+\hlf}^{\eta}}{\Delta t/2}+\frac{\hat{S}_{xx,\ell+1}^{\eta+\hlf}-\hat{S}_{xx,\ell}^{\eta+\hlf}}{\Delta x}\\
    -\frac{q}{m}\left(
    n_{\ell+\hlf}^{\eta+\hlf}E_{x,\ell+\hlf}^{\eta+\hlf}+\Gamma_{y,\ell+\hlf}^{\eta+\hlf}B_{z,\ell+\hlf}^{\eta+\hlf}-\Gamma_{z,\ell+\hlf}^{\eta+\hlf}B_{y,\ell+\hlf}^{\eta+\hlf}
    \right)
    -\gamma^{\eta+\hlf}_{\Gamma_x,\ell+\hlf}=0,
\end{multline}
\begin{multline}
    \frac{\Gamma_{y,\ell}^{\eta+\hlf}-\Gamma_{y,\ell}^{\eta}}{\Delta t/2}+\frac{\hat{S}_{xy,\ell+\hlf}^{\eta+\hlf}-\hat{S}_{xy,\ell-\hlf}^{\eta+\hlf}}{\Delta x}\\
    -\frac{q}{m}\left(
    n_{\ell}^{\eta+\hlf}E_{y,\ell}^{\eta+\hlf}+\Gamma_{z,\ell}^{\eta+\hlf}B_{x,\ell}^{\eta+\hlf}-\Gamma_{x,\ell}^{\eta+\hlf}B_{z,\ell}^{\eta+\hlf}
    \right)
    -\gamma^{\eta+\hlf}_{\Gamma_y,\ell}=0,
\end{multline}
\begin{multline}
    \frac{\Gamma_{z,\ell}^{\eta+\hlf}-\Gamma_{z,\ell}^{\eta}}{\Delta t/2}+\frac{\hat{S}_{xz,\ell+\hlf}^{\eta+\hlf}-\hat{S}_{xz,\ell-\hlf}^{\eta+\hlf}}{\Delta x}\\
    -\frac{q}{m}\left(
    n_{\ell}^{\eta+\hlf}E_{z,\ell}^{\eta+\hlf}+\Gamma_{x,\ell}^{\eta+\hlf}B_{y,\ell}^{\eta+\hlf}-\Gamma_{y,\ell}^{\eta+\hlf}B_{x,\ell}^{\eta+\hlf}
    \right)
    -\gamma^{\eta+\hlf}_{\Gamma_z,\ell}=0,
\end{multline}
\begin{equation}
    \epsilon_0\frac{E_{x,\ell+\hlf}^{\eta+1}-E_{x,\ell+\hlf}^{\eta}}{\Delta t}+
    \left(\sum_sq_s\Gamma_{s,x,\ell+\hlf}^{\eta+\hlf}-\frac{1}{N_x}\sum_{s,\ell+\hlf}q_s\Gamma_{s,x,\ell+\hlf}^{\eta+\hlf}\right)=0
\end{equation}
\begin{equation}
    \frac{A_{y,\ell+1}^{\eta+\hlf}-2A_{y,\ell}^{\eta+\hlf}+A_{y,\ell-1}^{\eta+\hlf}}{\Delta x^2}+\mu_0\left(\sum_sq_s\Gamma_{s,y,\ell}^{\eta+\hlf}-\frac{1}{N_x}\sum_{s,\ell}q_s\Gamma_{s,y,\ell}^{\eta+\hlf}\right)=0,
\end{equation}
\begin{equation}
    \frac{A_{z,\ell+1}^{\eta+\hlf}-2A_{z,\ell}^{\eta+\hlf}+A_{z,\ell-1}^{\eta+\hlf}}{\Delta x^2}+\mu_0\left(\sum_sq_s\Gamma_{s,z,\ell}^{\eta+\hlf}-\frac{1}{N_x}\sum_{s,\ell}q_s\Gamma_{s,z,\ell}^{\eta+\hlf}\right)=0,
\end{equation}
\begin{multline}
    \gamma^{\eta+\hlf}_{\Gamma_x,\ell+\hlf}=\frac{\Gamma_{x,\ell+\hlf}^{\eta+\hlf,\HO}-\Gamma_{x,\ell+\hlf}^{\eta,\HO}}{\Delta t/2}+\frac{\hat{S}_{xx,\ell+1}^{\eta+\hlf}-\hat{S}_{xx,\ell}^{\eta+\hlf}}{\Delta x}\\
    -\frac{q}{m}\left(
    n_{\ell+\hlf}^{\eta+\hlf,\HO}E_{x,\ell+\hlf}^{\eta+\hlf}+\Gamma_{y,\ell+\hlf}^{\eta+\hlf,\HO}B_{z,\ell+\hlf}^{\eta+\hlf}-\Gamma_{z,\ell+\hlf}^{\eta+\hlf,\HO}B_{y,\ell+\hlf}^{\eta+\hlf}
    \right),
\end{multline}
\begin{multline}
    \gamma^{\eta+\hlf}_{\Gamma_y,\ell}=
    \frac{\Gamma_{y,\ell}^{\eta+\hlf,\HO}-\Gamma_{y,\ell}^{\eta,\HO}}{\Delta t/2}+\frac{\hat{S}_{xy,\ell+\hlf}^{\eta+\hlf}-\hat{S}_{xy,\ell-\hlf}^{\eta+\hlf}}{\Delta x}\\
    -\frac{q}{m}\left(
    n_{\ell}^{\eta+\hlf,\HO}E_{y,\ell}^{\eta+\hlf}+\Gamma_{z,\ell}^{\eta+\hlf,\HO}B_{x,\ell}^{\eta+\hlf}-\Gamma_{x,\ell}^{\eta+\hlf,\HO}B_{z,\ell}^{\eta+\hlf}
    \right)
    ,
\end{multline}
\begin{multline}
    \gamma^{\eta+\hlf}_{\Gamma_z,\ell}=
    \frac{\Gamma_{z,\ell}^{\eta+\hlf,\HO}-\Gamma_{z,\ell}^{\eta,\HO}}{\Delta t/2}+\frac{\hat{S}_{xz,\ell+\hlf}^{\eta+\hlf}-\hat{S}_{xz,\ell-\hlf}^{\eta+\hlf}}{\Delta x}\\
    -\frac{q}{m}\left(
    n_{\ell}^{\eta+\hlf,\HO}E_{z,\ell}^{\eta+\hlf}+\Gamma_{x,\ell}^{\eta+\hlf,\HO}B_{y,\ell}^{\eta+\hlf}-\Gamma_{y,\ell}^{\eta+\hlf,\HO}B_{x,\ell}^{\eta+\hlf}
    \right)
    .
\end{multline}
where
\begin{equation}
    \textrm{Conservative}:
    \hat{S}_{ij} = n\tilde{S}_{ij}^{\HO}
    \hspace{25pt}
    \textrm{Primitive}:
    \hat{S}_{ij} = n\tilde{T}_{ij}^{\HO}+\frac{\Gamma_i\Gamma_{j}}{n}.
\end{equation}
\subsection{5-moment system}
\begin{equation}\label{eq:5M_n}
    \frac{n^{\eta+1}_\ell-n^{\eta}_\ell}{\Delta t}+\frac{\Gamma_{x,\ell+\hlf}^{\eta+\hlf}-\Gamma_{x,\ell-\hlf}^{\eta+\hlf}}{\Delta x}=0,
\end{equation}
\begin{multline}
    \frac{\Gamma_{x,\ell+\hlf}^{\eta+\hlf}-\Gamma_{x,\ell+\hlf}^{\eta}}{\Delta t/2}+\frac{{S}_{xx,\ell+1}^{\eta+\hlf}-{S}_{xx,\ell}^{\eta+\hlf}}{\Delta x}\\
    -\frac{q}{m}\left(
    n_{\ell+\hlf}^{\eta+\hlf}E_{x,\ell+\hlf}^{\eta+\hlf}+\Gamma_{y,\ell+\hlf}^{\eta+\hlf}B_{z,\ell+\hlf}^{\eta+\hlf}-\Gamma_{z,\ell+\hlf}^{\eta+\hlf}B_{y,\ell+\hlf}^{\eta+\hlf}
    \right)
    -\gamma^{\eta+\hlf}_{\Gamma_x,\ell+\hlf}=0,
\end{multline}
\begin{multline}
    \frac{\Gamma_{y,\ell}^{\eta+\hlf}-\Gamma_{y,\ell}^{\eta}}{\Delta t/2}+\frac{\hat{S}_{xy,\ell+\hlf}^{\eta+\hlf}-\hat{S}_{xy,\ell-\hlf}^{\eta+\hlf}}{\Delta x}\\
    -\frac{q}{m}\left(
    n_{\ell}^{\eta+\hlf}E_{y,\ell}^{\eta+\hlf}+\Gamma_{z,\ell}^{\eta+\hlf}B_{x,\ell}^{\eta+\hlf}-\Gamma_{x,\ell}^{\eta+\hlf}B_{z,\ell}^{\eta+\hlf}
    \right)
    -\gamma^{\eta+\hlf}_{\Gamma_y,\ell}=0,
\end{multline}
\begin{multline}
    \frac{\Gamma_{z,\ell}^{\eta+\hlf}-\Gamma_{z,\ell}^{\eta}}{\Delta t/2}+\frac{\hat{S}_{xz,\ell+\hlf}^{\eta+\hlf}-\hat{S}_{xz,\ell-\hlf}^{\eta+\hlf}}{\Delta x}\\
    -\frac{q}{m}\left(
    n_{\ell}^{\eta+\hlf}E_{z,\ell}^{\eta+\hlf}+\Gamma_{x,\ell}^{\eta+\hlf}B_{y,\ell}^{\eta+\hlf}-\Gamma_{y,\ell}^{\eta+\hlf}B_{x,\ell}^{\eta+\hlf}
    \right)
    -\gamma^{\eta+\hlf}_{\Gamma_z,\ell}=0,
\end{multline}
\begin{multline}\label{eq:5M_Sxx}
    \frac{S_{xx,\ell}^{\eta+1}-S_{xx,\ell}^{\eta}}{\Delta t}+\frac{\hat{Q}_{xxx,\ell+\hlf}^{\eta+\hlf}-\hat{Q}_{xxx,\ell-\hlf}^{\eta+\hlf}}{\Delta x}\\
    -\frac{2q}{m}\left(
    \Gamma_{x,\ell}^{\eta+\hlf}E_{x,\ell}^{\eta+\hlf}
    +\hat{S}_{xy,\ell}^{\eta+\hlf}B_{z,\ell}^{\eta+\hlf}
    -\hat{S}_{xz,\ell}^{\eta+\hlf}B_{y,\ell}^{\eta+\hlf}
    \right)
    -\gamma^{\eta+\hlf}_{S_{xx},\ell}=0,
\end{multline}
\begin{equation}
    \epsilon_0\frac{E_{x,\ell+\hlf}^{\eta+1}-E_{x,\ell+\hlf}^{\eta}}{\Delta t}+
    \left(\sum_sq_s\Gamma_{s,x,\ell+\hlf}^{\eta+\hlf}-\frac{1}{N_x}\sum_{s,\ell+\hlf}q_s\Gamma_{s,x,\ell+\hlf}^{\eta+\hlf}\right)=0
\end{equation}
\begin{equation}
    \frac{A_{y,\ell+1}^{\eta+\hlf}-2A_{y,\ell}^{\eta+\hlf}+A_{y,\ell-1}^{\eta+\hlf}}{\Delta x^2}+\mu_0\left(\sum_sq_s\Gamma_{s,y,\ell}^{\eta+\hlf}-\frac{1}{N_x}\sum_{s,\ell}q_s\Gamma_{s,y,\ell}^{\eta+\hlf}\right)=0,
\end{equation}
\begin{equation}
    \frac{A_{z,\ell+1}^{\eta+\hlf}-2A_{z,\ell}^{\eta+\hlf}+A_{z,\ell-1}^{\eta+\hlf}}{\Delta x^2}+\mu_0\left(\sum_sq_s\Gamma_{s,z,\ell}^{\eta+\hlf}-\frac{1}{N_x}\sum_{s,\ell}q_s\Gamma_{s,z,\ell}^{\eta+\hlf}\right)=0,
\end{equation}
\begin{multline}
    \gamma^{\eta+\hlf}_{\Gamma_x,\ell+\hlf}=\frac{\Gamma_{x,\ell+\hlf}^{\eta+\hlf,\HO}-\Gamma_{x,\ell+\hlf}^{\eta,\HO}}{\Delta t/2}+\frac{\hat{S}_{xx,\ell+1}^{\eta+\hlf}-\hat{S}_{xx,\ell}^{\eta+\hlf}}{\Delta x}\\
    -\frac{q}{m}\left(
    n_{\ell+\hlf}^{\eta+\hlf,\HO}E_{x,\ell+\hlf}^{\eta+\hlf}+\Gamma_{y,\ell+\hlf}^{\eta+\hlf,\HO}B_{z,\ell+\hlf}^{\eta+\hlf}-\Gamma_{z,\ell+\hlf}^{\eta+\hlf,\HO}B_{y,\ell+\hlf}^{\eta+\hlf}
    \right),
\end{multline}
\begin{multline}
    \gamma^{\eta+\hlf}_{\Gamma_y,\ell}=
    \frac{\Gamma_{y,\ell}^{\eta+\hlf,\HO}-\Gamma_{y,\ell}^{\eta,\HO}}{\Delta t/2}+\frac{\hat{S}_{xy,\ell+\hlf}^{\eta+\hlf}-\hat{S}_{xy,\ell-\hlf}^{\eta+\hlf}}{\Delta x}\\
    -\frac{q}{m}\left(
    n_{\ell}^{\eta+\hlf,\HO}E_{y,\ell}^{\eta+\hlf}+\Gamma_{z,\ell}^{\eta+\hlf,\HO}B_{x,\ell}^{\eta+\hlf}-\Gamma_{x,\ell}^{\eta+\hlf,\HO}B_{z,\ell}^{\eta+\hlf}
    \right)
    ,
\end{multline}
\begin{multline}
    \gamma^{\eta+\hlf}_{\Gamma_z,\ell}=
    \frac{\Gamma_{z,\ell}^{\eta+\hlf,\HO}-\Gamma_{z,\ell}^{\eta,\HO}}{\Delta t/2}+\frac{\hat{S}_{xz,\ell+\hlf}^{\eta+\hlf}-\hat{S}_{xz,\ell-\hlf}^{\eta+\hlf}}{\Delta x}\\
    -\frac{q}{m}\left(
    n_{\ell}^{\eta+\hlf,\HO}E_{z,\ell}^{\eta+\hlf}+\Gamma_{x,\ell}^{\eta+\hlf,\HO}B_{y,\ell}^{\eta+\hlf}-\Gamma_{y,\ell}^{\eta+\hlf,\HO}B_{x,\ell}^{\eta+\hlf}
    \right)
    .
\end{multline}
\begin{multline}
    \gamma^{\eta+\hlf}_{S_{xx},\ell}=\frac{S_{xx,\ell}^{\eta+1,\HO}-S_{xx,\ell}^{\eta,\HO}}{\Delta t}+\frac{\hat{Q}_{xxx,\ell+\hlf}^{\eta+\hlf}-\hat{Q}_{xxx,\ell-\hlf}^{\eta+\hlf}}{\Delta x}\\
    -\frac{2q}{m}\left(
    \Gamma_{x,\ell}^{\eta+\hlf,\HO}E_{x,\ell}^{\eta+\hlf}
    +\hat{S}_{xy,\ell}^{\eta+\hlf}B_{z,\ell}^{\eta+\hlf}
    -\hat{S}_{xz,\ell}^{\eta+\hlf}B_{y,\ell}^{\eta+\hlf}
    \right)
    ,
\end{multline}
where
\begin{equation}
    \textrm{Conservative}:
    \hat{Q}_{xxx} = n\tilde{Q}^{\HO}_{xxx}
    \hspace{25pt}
    \textrm{Primitive}:
    \hat{Q}_{xxi} = n\tilde{q}^{\HO}_{xxi}+\frac{2\Gamma_xS_{xi}+2\Gamma_iS_{xx}}{n}-2\frac{\Gamma_i\Gamma_x^2}{n^2}.
\end{equation}
and $\hat{S}$ is defined as above.
\subsection{7-moment system}
\begin{equation}\label{eq:7M_n}
    \frac{n^{\eta+1}_\ell-n^{\eta}_\ell}{\Delta t}+\frac{\Gamma_{x,\ell+\hlf}^{\eta+\hlf}-\Gamma_{x,\ell-\hlf}^{\eta+\hlf}}{\Delta x}=0,
\end{equation}
\begin{multline}
    \frac{\Gamma_{x,\ell+\hlf}^{\eta+\hlf}-\Gamma_{x,\ell+\hlf}^{\eta}}{\Delta t/2}+\frac{{S}_{xx,\ell+1}^{\eta+\hlf}-{S}_{xx,\ell}^{\eta+\hlf}}{\Delta x}\\
    -\frac{q}{m}\left(
    n_{\ell+\hlf}^{\eta+\hlf}E_{x,\ell+\hlf}^{\eta+\hlf}+\Gamma_{y,\ell+\hlf}^{\eta+\hlf}B_{z,\ell+\hlf}^{\eta+\hlf}-\Gamma_{z,\ell+\hlf}^{\eta+\hlf}B_{y,\ell+\hlf}^{\eta+\hlf}
    \right)
    -\gamma^{\eta+\hlf}_{\Gamma_x,\ell+\hlf}=0,
\end{multline}
\begin{multline}
    \frac{\Gamma_{y,\ell}^{\eta+\hlf}-\Gamma_{y,\ell}^{\eta}}{\Delta t/2}+\frac{{S}_{xy,\ell+\hlf}^{\eta+\hlf}-{S}_{xy,\ell-\hlf}^{\eta+\hlf}}{\Delta x}\\
    -\frac{q}{m}\left(
    n_{\ell}^{\eta+\hlf}E_{y,\ell}^{\eta+\hlf}+\Gamma_{z,\ell}^{\eta+\hlf}B_{x,\ell}^{\eta+\hlf}-\Gamma_{x,\ell}^{\eta+\hlf}B_{z,\ell}^{\eta+\hlf}
    \right)
    -\gamma^{\eta+\hlf}_{\Gamma_y,\ell}=0,
\end{multline}
\begin{multline}
    \frac{\Gamma_{z,\ell}^{\eta+\hlf}-\Gamma_{z,\ell}^{\eta}}{\Delta t/2}+\frac{{S}_{xz,\ell+\hlf}^{\eta+\hlf}-{S}_{xz,\ell-\hlf}^{\eta+\hlf}}{\Delta x}\\
    -\frac{q}{m}\left(
    n_{\ell}^{\eta+\hlf}E_{z,\ell}^{\eta+\hlf}+\Gamma_{x,\ell}^{\eta+\hlf}B_{y,\ell}^{\eta+\hlf}-\Gamma_{y,\ell}^{\eta+\hlf}B_{x,\ell}^{\eta+\hlf}
    \right)
    -\gamma^{\eta+\hlf}_{\Gamma_z,\ell}=0,
\end{multline}
\begin{multline}\label{eq:7M_Sxx}
    \frac{S_{xx,\ell}^{\eta+1}-S_{xx,\ell}^{\eta}}{\Delta t}+\frac{\hat{Q}_{xxx,\ell+\hlf}^{\eta+\hlf}-\hat{Q}_{xxx,\ell-\hlf}^{\eta+\hlf}}{\Delta x}\\
    -\frac{2q}{m}\left(
    \Gamma_{x,\ell}^{\eta+\hlf}E_{x,\ell}^{\eta+\hlf}
    +S_{xy,\ell}^{\eta+\hlf}B_{z,\ell}^{\eta+\hlf}
    -{S}_{xz,\ell}^{\eta+\hlf}B_{y,\ell}^{\eta+\hlf}
    \right)
    -\gamma^{\eta+\hlf}_{S_{xx},\ell}=0,
\end{multline}
\begin{multline}\label{eq:7M_Sxy}
    \frac{S_{xy,\ell+\hlf}^{\eta+1}-S_{xy,\ell+\hlf}^{\eta}}{\Delta t}+\frac{\hat{Q}_{xxy,\ell+1}^{\eta+\hlf}-\hat{Q}_{xxy,\ell}^{\eta+\hlf}}{\Delta x}
    -\frac{q}{m}\Bigg[
    \Gamma_{x,\ell}^{\eta+\hlf}E_{y,\ell}^{\eta+\hlf}
    +\Gamma_{y,\ell}^{\eta+\hlf}E_{x,\ell}^{\eta+\hlf}\\
    +S_{xz,\ell}^{\eta+\hlf}B_{x,\ell}^{\eta+\hlf}
    +\left(\hat{S}_{yy,\ell}^{\eta+\hlf}-{S}_{xx,\ell}^{\eta+\hlf}\right)B_{z,\ell}^{\eta+\hlf}
    -\hat{S}_{yz,\ell}^{\eta+\hlf}B_{y,\ell}^{\eta+\hlf}
    \Bigg]
    -\gamma^{\eta+\hlf}_{S_{xy},\ell+\hlf}=0,
\end{multline}
\begin{multline}\label{eq:7M_Sxz}
    \frac{S_{xz,\ell+\hlf}^{\eta+1}-S_{xz,\ell+\hlf}^{\eta}}{\Delta t}+\frac{\hat{Q}_{xxz,\ell+1}^{\eta+\hlf}-\hat{Q}_{xxz,\ell}^{\eta+\hlf}}{\Delta x}
    -\frac{q}{m}\Bigg[
    \Gamma_{x,\ell}^{\eta+\hlf}E_{z,\ell}^{\eta+\hlf}
    +\Gamma_{z,\ell}^{\eta+\hlf}E_{x,\ell}^{\eta+\hlf}\\
    +\hat{S}_{yz,\ell}^{\eta+\hlf}B_{z,\ell}^{\eta+\hlf}
    +\left({S}_{xx,\ell}^{\eta+\hlf}-\hat{S}_{zz,\ell}^{\eta+\hlf}\right)B_{y,\ell}^{\eta+\hlf}
    -{S}_{xy,\ell}^{\eta+\hlf}B_{x,\ell}^{\eta+\hlf}
    \Bigg]
    -\gamma^{\eta+\hlf}_{S_{xz},\ell+\hlf}=0,
\end{multline}
\begin{equation}
    \epsilon_0\frac{E_{x,\ell+\hlf}^{\eta+1}-E_{x,\ell+\hlf}^{\eta}}{\Delta t}+
    \left(\sum_sq_s\Gamma_{s,x,\ell+\hlf}^{\eta+\hlf}-\frac{1}{N_x}\sum_{s,\ell+\hlf}q_s\Gamma_{s,x,\ell+\hlf}^{\eta+\hlf}\right)=0
\end{equation}
\begin{equation}
    \frac{A_{y,\ell+1}^{\eta+\hlf}-2A_{y,\ell}^{\eta+\hlf}+A_{y,\ell-1}^{\eta+\hlf}}{\Delta x^2}+\mu_0\left(\sum_sq_s\Gamma_{s,y,\ell}^{\eta+\hlf}-\frac{1}{N_x}\sum_{s,\ell}q_s\Gamma_{s,y,\ell}^{\eta+\hlf}\right)=0,
\end{equation}
\begin{equation}
    \frac{A_{z,\ell+1}^{\eta+\hlf}-2A_{z,\ell}^{\eta+\hlf}+A_{z,\ell-1}^{\eta+\hlf}}{\Delta x^2}+\mu_0\left(\sum_sq_s\Gamma_{s,z,\ell}^{\eta+\hlf}-\frac{1}{N_x}\sum_{s,\ell}q_s\Gamma_{s,z,\ell}^{\eta+\hlf}\right)=0,
\end{equation}
\begin{multline}
    \gamma^{\eta+\hlf}_{\Gamma_x,\ell+\hlf}=\frac{\Gamma_{x,\ell+\hlf}^{\eta+\hlf,\HO}-\Gamma_{x,\ell+\hlf}^{\eta,\HO}}{\Delta t/2}+\frac{n_{\ell+1}^{\eta+\hlf,\HO}\hat{S}_{xx,\ell+1}^{\eta+\hlf}-n_{\ell}^{\eta+\hlf,\HO}\hat{S}_{xx,\ell}^{\eta+\hlf}}{\Delta x}\\
    -\frac{q}{m}\left(
    n_{\ell+\hlf}^{\eta+\hlf,\HO}E_{x,\ell+\hlf}^{\eta+\hlf}+\Gamma_{y,\ell+\hlf}^{\eta+\hlf,\HO}B_{z,\ell+\hlf}^{\eta+\hlf}-\Gamma_{z,\ell+\hlf}^{\eta+\hlf,\HO}B_{y,\ell+\hlf}^{\eta+\hlf}
    \right),
\end{multline}
\begin{multline}
    \gamma^{\eta+\hlf}_{\Gamma_y,\ell}=
    \frac{\Gamma_{y,\ell}^{\eta+\hlf,\HO}-\Gamma_{y,\ell}^{\eta,\HO}}{\Delta t/2}+\frac{n_{\ell+\hlf}^{\eta+\hlf}\hat{S}_{xy,\ell+\hlf}^{\eta+\hlf}-n_{\ell-\hlf}^{\eta+\hlf}\hat{S}_{xy,\ell-\hlf}^{\eta+\hlf}}{\Delta x}\\
    -\frac{q}{m}\left(
    n_{\ell}^{\eta+\hlf,\HO}E_{y,\ell}^{\eta+\hlf}+\Gamma_{z,\ell}^{\eta+\hlf,\HO}B_{x,\ell}^{\eta+\hlf}-\Gamma_{x,\ell}^{\eta+\hlf,\HO}B_{z,\ell}^{\eta+\hlf}
    \right)
    ,
\end{multline}
\begin{multline}
    \gamma^{\eta+\hlf}_{\Gamma_z,\ell}=
    \frac{\Gamma_{z,\ell}^{\eta+\hlf,\HO}-\Gamma_{z,\ell}^{\eta,\HO}}{\Delta t/2}+\frac{n_{\ell+\hlf}^{\eta+\hlf,\HO}\hat{S}_{xz,\ell+\hlf}^{\eta+\hlf}-n_{\ell-\hlf}^{\eta+\hlf,\HO}\hat{S}_{xz,\ell-\hlf}^{\eta+\hlf}}{\Delta x}\\
    -\frac{q}{m}\left(
    n_{\ell}^{\eta+\hlf,\HO}E_{z,\ell}^{\eta+\hlf}+\Gamma_{x,\ell}^{\eta+\hlf,\HO}B_{y,\ell}^{\eta+\hlf}-\Gamma_{y,\ell}^{\eta+\hlf,\HO}B_{x,\ell}^{\eta+\hlf}
    \right)
    .
\end{multline}
\begin{multline}
    \gamma^{\eta+\hlf}_{S_{xx},\ell}=\frac{S_{xx,\ell}^{\eta+1,\HO}-S_{xx,\ell}^{\eta,\HO}}{\Delta t}+\frac{\hat{Q}_{xxx,\ell+\hlf}^{\eta+\hlf}-\hat{Q}_{xxx,\ell-\hlf}^{\eta+\hlf}}{\Delta x}\\
    -\frac{2q}{m}\left(
    \Gamma_{x,\ell}^{\eta+\hlf,\HO}E_{x,\ell}^{\eta+\hlf}
    +{S}_{xy,\ell}^{\eta+\hlf,\HO}B_{z,\ell}^{\eta+\hlf}
    -{S}_{xz,\ell}^{\eta+\hlf,\HO}B_{y,\ell}^{\eta+\hlf}
    \right)
    ,
\end{multline}
\begin{multline}
    \gamma^{\eta+\hlf}_{S_{xy},\ell}=\frac{S_{xy,\ell+\hlf}^{\eta+1,\HO}-S_{xy,\ell+\hlf}^{\eta,\HO}}{\Delta t}+\frac{\hat{Q}_{xxy,\ell+1}^{\eta+\hlf}-\hat{Q}_{xxy,\ell}^{\eta+\hlf}}{\Delta x}
    -\frac{q}{m}\Bigg[
    \Gamma_{x,\ell}^{\eta+\hlf,\HO}E_{y,\ell}^{\eta+\hlf}
    +\Gamma_{y,\ell}^{\eta+\hlf,\HO}E_{x,\ell}^{\eta+\hlf}\\
    +S_{xz,\ell}^{\eta+\hlf,\HO}B_{x,\ell}^{\eta+\hlf}
    +\left(\hat{S}_{yy,\ell}^{\eta+\hlf}-{S}_{xx,\ell}^{\eta+\hlf,\HO}\right)B_{z,\ell}^{\eta+\hlf}
    -\hat{S}_{yz,\ell}^{\eta+\hlf,\HO}B_{y,\ell}^{\eta+\hlf}
    \Bigg]
    ,
\end{multline}
\begin{multline}
    \gamma^{\eta+\hlf}_{S_{xz},\ell}=\frac{S_{xz,\ell+\hlf}^{\eta+1,\HO}-S_{xz,\ell+\hlf}^{\eta,\HO}}{\Delta t}+\frac{\hat{Q}_{xxz,\ell+1}^{\eta+\hlf}-\hat{Q}_{xxz,\ell}^{\eta+\hlf}}{\Delta x}
    -\frac{q}{m}\Bigg[
    \Gamma_{x,\ell}^{\eta+\hlf,\HO}E_{z,\ell}^{\eta+\hlf}
    +\Gamma_{z,\ell}^{\eta+\hlf,\HO}E_{x,\ell}^{\eta+\hlf}\\
    +\hat{S}_{yz,\ell}^{\eta+\hlf,\HO}B_{z,\ell}^{\eta+\hlf}
    +\left({S}_{xx,\ell}^{\eta+\hlf,\HO}-\hat{S}_{zz,\ell}^{\eta+\hlf,\HO}\right)B_{y,\ell}^{\eta+\hlf}
    -{S}_{xy,\ell}^{\eta+\hlf,\HO}B_{x,\ell}^{\eta+\hlf}
    \Bigg],
\end{multline}
where $\hat{S}_{ij}$ and $\hat{Q}_{xxi}$ are defined as above.
\vspace{10pt}

\section{Anderson mixing} \label{sec:AppAA}

The Anderson mixing scheme \cite{anderson1965iterative,walker2011anderson,willert2014leveraging} updates the estimate of the solution vector $\mathbf{U}^{(y+1)}$ in the solution of Eqn.~\eqref{eq:HOLO_res} by applying an algorithm at each iteration that takes into account a maximum of $h$ histories of previous residuals and solution vectors. The Anderson mixing algorithm is presented in Algorithm~\ref{alg:Anderson}.\par 
Recall that $y$ is the HOLO iteration index so $h_y=\min(h,y)$ is the maximum number of histories that can be used and $\textbf{r}^{(y)}$ is the HOLO residual, i.e., the difference in $\textbf{U}$ between HOLO iterations. $\vec{\kappa}$ is a weight vector that defines the linear combination of $\textbf{U}^{(y)}$ used to calculate $\textbf{U}^{(y+1)}$ and $\vec{\alpha}$ is a dummy variable used to calculate the optimal $\vec{\kappa}$.

\begin{algorithm}
    \caption{Anderson Mixing Algorithm. 
    }\label{alg:Anderson}
    \begin{algorithmic}
        \Function{AndersonMixing}{$\textbf{r},\textbf{U},y,h$}
        \State $h_y\gets\min(h,y)$
        \State $\mathbf{\mathcal{R}}_y \gets \left( \textbf{r}^{(y-h_y)},\ldots,\textbf{r}^{(y)}\right)$
        \State Compute $\vec{\kappa} = \left(\kappa^{(y)}_0,\ldots,\kappa^{(y)}_{h_y}\right)^T$, where
        \vspace{-3ex}
                \begin{equation*}
                    \vec{\kappa} = \textrm{arg}\min_{\vec{\alpha}}\parallel \mathbf{\mathcal{R}}_y\vec{\alpha}\parallel_2\hspace{15pt} \textrm{subject to} \hspace{15pt}
                    \sum_{\xi=0}^{h_y}\kappa_\xi=1.
                    \vspace{-3ex}
                \end{equation*}
        \State $\textbf{U}^{(y+1)} \gets\sum_{\xi=0}^{h_y}\kappa_\xi\textbf{U}^{(y-\xi)}$
        \State \textbf{return} $\textbf{U}^{(y+1)}$
        \EndFunction
    \end{algorithmic}
\end{algorithm}

\vspace{10pt}
\pagebreak
\bibliography{sample}{}
\end{document}